\documentclass[aps, amsmath, amssymb, nofootinbib, prd, reprint, superscriptaddress]{revtex4-2}
 
\usepackage[x11names]{xcolor}
\usepackage[colorlinks=true, linkcolor=RoyalBlue4, citecolor=RoyalBlue4, urlcolor=RoyalBlue4]{hyperref}
\usepackage[capitalize]{cleveref}
\usepackage{booktabs}
\usepackage{graphicx}
\usepackage{mathtools}
\usepackage{orcidlink}
\usepackage{siunitx}

\DeclareSIUnit{\hubble}{\textit{h}}
\DeclareSIUnit{\parsec}{pc}

\begin{document}

\title{Less isn't more: Cosmological bounds on the neutrino masses\\
are robust to changes in the neutrino abundance}

\author{Sebastian J. Figueroa\,\orcidlink{0009-0009-3155-611X}}
\email{sebastian.figueroa@unc.edu}
\affiliation{Department of Physics and Astronomy, University of North Carolina at Chapel Hill,\\
Chapel Hill, North Carolina 27599, USA}

\author{Alexander C. Sobotka\,\orcidlink{0000-0002-7576-5417}}
\email{asobotka@highpoint.edu}
\affiliation{Department of Physics, High Point University,\\
High Point, North Carolina 27268, USA}

\author{Adrienne L. Erickcek\,\orcidlink{0000-0002-0901-3591}}
\email{erickcek@physics.unc.edu}
\affiliation{Department of Physics and Astronomy, University of North Carolina at Chapel Hill,\\
Chapel Hill, North Carolina 27599, USA}

\date{September 14, 2026}

\begin{abstract}
We investigate how neutrino-mass constraints from cosmology depend on the assumed thermal history of the universe. Photon injection after neutrino decoupling would decrease the neutrino abundance inferred from the temperature of the cosmic microwave background~(CMB), potentially loosening the upper limit on their masses. We first evaluate how the cosmological neutrino-mass bound is altered by the decay of massive particles into photons and dark radiation after Big Bang nucleosynthesis. To test the robustness of this constraint more generally, we also assess the impact of varying the temperature of the cosmic neutrino background without assuming a specific physical mechanism. We perform Markov chain Monte Carlo analyses of both frameworks with primary CMB observations from \textit{Planck}, CMB lensing measurements from \textit{Planck} and the Atacama Cosmology Telescope, and baryon acoustic oscillation data from the Dark Energy Spectroscopic Instrument. For the degenerate mass ordering, the 95\% credible limit tightens from $\sum m_{\nu} < \qty{0.0691}{\electronvolt}$ in a standard thermal history to $\sum m_{\nu} < \qty{0.0652}{\electronvolt}$ if the massive particles decay exclusively into photons, while the injection of dark radiation in addition to photons slightly relaxes this limit to $\sum m_{\nu} < \qty{0.0710}{\electronvolt}$. The same pattern holds for the normal and inverted orderings, and the decay scenario shifts the bound on the sum of the neutrino masses by at most $\qty{0.004}{\electronvolt}$ for fixed mass splittings. Allowing model-agnostic changes in the neutrino-to-photon ratio yields a 95\% credible limit of $\sum m_{\nu} < \qty{0.0724}{\electronvolt}$ for the degenerate ordering, indicating that the stringency of our neutrino-mass bounds is not driven by constraints on the decay scenario. We find that the neutrino temperature and the sum of the neutrino masses are positively correlated, which implies that reducing the pre-recombination radiation density will only worsen the emerging tension between cosmological bounds on the neutrino masses and the measured mass splittings.
\end{abstract}

\maketitle

\section{Introduction}
\label{sec:introduction}

The discovery of neutrino oscillations revealed that at least two of the three neutrino species possess nonzero mass~\cite{Super-Kamiokande:1998kpq, SNO:2002tuh}. Oscillation experiments do not directly measure the absolute neutrino masses, but instead constrain two mass-squared splittings. The sign of the solar splitting is fixed by the Mikheyev--Smirnov--Wolfenstein effect~\cite{Langacker:1986jv}, whereas the sign of the much larger atmospheric splitting is currently unknown. Two mass orderings are therefore possible, with $m_1 < m_2 \ll m_3$ in the normal ordering~(NO) and $m_3 \ll m_1 < m_2$ in the inverted ordering~(IO). If the lightest neutrino is assumed to be massless in either ordering, the measured solar and atmospheric splittings yield one-sided 95\% confidence intervals~(CI) on the sum of the neutrino masses of $\sum m_{\nu} \geq \qty{0.058}{\electronvolt}$ for the NO and $\sum m_{\nu} \geq \qty{0.098}{\electronvolt}$ for the IO~\cite{Esteban:2024eli}.

Massive neutrinos affect cosmological observables by altering the expansion history and the growth of small-scale structure. Since they contribute to the radiation density at matter--radiation~(MR) equality while constituting part of the matter density today, neutrinos have a unique effect on the expansion history of the universe. Neutrinos also affect structure growth by free streaming out of overdense regions, erasing their perturbations on scales smaller than the comoving distance they travel. As neutrinos do not efficiently source gravitational potentials on these small scales, they further suppress structure formation by slowing the growth of cold dark matter~(CDM) and baryon perturbations~\cite{Wong:2011ip, Lattanzi:2017ubx, Bertolez-Martinez:2024wez}. 

Observations of the primary cosmic microwave background~(CMB) anisotropies through the TT, TE, and EE power spectra constrain the distance to the surface of last scattering, which depends on the neutrino masses and $H_0$, the present-day expansion rate. The degeneracy between these two parameters is partially lifted by the gravitational lensing of the CMB, which is sensitive to the amplitude of matter clustering on the scales suppressed by neutrino free streaming. Bounds on the neutrino masses are further strengthened by measurements of the baryon acoustic oscillations~(BAO), which constrain the late-time expansion history and break geometric degeneracies in analyses limited to CMB data~\cite{Archidiacono:2016lnv, Vagnozzi:2017ovm, Lynch:2025ine}.

In its recent DR2 analysis, the Dark Energy Spectroscopic Instrument~(DESI) Collaboration obtained one of the strongest cosmological constraints on the neutrino masses by combining DESI BAO data with primary CMB spectra from \textit{Planck} and CMB lensing power spectra from \textit{Planck} and the Atacama Cosmology Telescope~(ACT). For the degenerate ordering~(DO), which divides the sum of the neutrino masses equally among the three mass eigenstates, this combination of CMB and BAO datasets yielded a one-sided 95\% credible interval~(CrI)\footnote{For both one-sided and two-sided constraints, we denote frequentist confidence intervals with CI and Bayesian credible intervals with CrI, reflecting their distinct statistical interpretations~\cite{Trotta:2008qt}.} of $\sum m_{\nu} < \qty{0.0642}{\electronvolt}$~(DO)~\cite{Elbers:2025vlz}. Comparing cosmological limits with the minimum values inferred from oscillation experiments reveals an increasingly narrow range of neutrino masses consistent with both sets of bounds~\cite{Gariazzo:2023joe, Jiang:2024viw}. Motivated by this tension, the DESI Collaboration expanded on the DR2 analysis with an effective neutrino mass parameter allowed to take negative values~\cite{Elbers:2024sha}, yielding a posterior that peaked in the negative-mass region and a corresponding constraint of $\sum m_{\nu, \, \mathrm{eff}} = \qty{-0.101(0.047:0.056)}{\electronvolt}$~(68\%~CrI)~\cite{Elbers:2025vlz}.

This pull toward unphysical negative neutrino masses may be a diagnostic of dataset or likelihood systematics, particularly those associated with the CMB lensing anomaly and the emerging discrepancy between different probes of the present-day matter density~\cite{Green:2024xbb, Cozzumbo:2025ewt}. Alternatively, the preference for negative neutrino masses may signal missing physics, including enhanced CMB lensing, modified structure growth, or deviations from the $\Lambda$CDM expansion history~\cite{Craig:2024tky, Graham:2025dqn}. Various models have been proposed to address this neutrino-mass tension, including dark forces~\cite{Graham:2025dqn, Esteban:2021ozz, Costa:2025kwt}, modified recombination~\cite{Sekiguchi:2020igz, Baryakhtar:2024rky, Lynch:2024hzh}, a larger optical depth to reionization~\cite{Jhaveri:2025neg, Sailer:2025lxj}, dynamical dark energy~\cite{Hannestad:2005gj, Lorenz:2017fgo, Vagnozzi:2018jhn, Upadhye:2017hdl, Sharma:2022ifr}, decaying dark matter~\cite{Poulin:2016nat, Montandon:2026vuc}, nonstandard neutrino interactions~\cite{Beacom:2004yd, Farzan:2015pca, Esteban:2022rjk, Das:2025asx}, neutrino decays~\cite{Serpico:2007pt, Chacko:2019nej, Escudero:2020ped, FrancoAbellan:2026ori}, and time-varying neutrino masses~\cite{Lorenz:2021alz, Huang:2022wmz, daFonseca:2023ury, Sen:2024pgb, Ghedini:2025epp}. 

The mounting tension between various cosmological probes makes assessing the robustness of neutrino-mass constraints increasingly important\footnote{Direct kinematic measurements of the tritium beta-decay endpoint provide a complementary constraint on the absolute neutrino mass scale. The Karlsruhe Tritium Neutrino Collaboration reports the leading upper limit on the effective electron antineutrino mass, $m_{\bar{\nu}_e} < \qty{0.45}{\electronvolt}$~(90\%~CI), which implies a bound on the sum of the neutrino masses of $\sum m_{\nu} < \qty{1.35}{\electronvolt}$~(90\%~CI)~\cite{KATRIN:2024cdt}.}~\cite{diValentino:2022njd, Naredo-Tuero:2024sgf, Shao:2024mag, RoyChoudhury:2025dhe, Chebat:2025kes, Hergt:2026moc, Qu:2026fkt}. In this work, we examine how a modified thermal history affects the inferred cosmological bound on the neutrino masses. The cosmic neutrino background~(C$\nu$B) has not been directly observed, so the relic neutrino density is inferred from its predicted ratio to the photon density. We first consider a massive hidden-sector particle, denoted by $Y$, that decays into photons and dark radiation~(DR) between Big Bang nucleosynthesis~(BBN) and recombination. Measurements of the primordial deuterium abundance limit the pre-decay contribution of $Y$ particles to $2.35\%$~(95\%~CrI) of the total energy density of the universe, assuming one massive neutrino species with $m_{\nu} = \qty{0.06}{\electronvolt}$~\cite{Sobotka:2022vrr}. This constraint was obtained by marginalizing over the fraction of $Y$ particles that decay into photons, but if applied to decays that produce photons exclusively, it would limit the reduction in the neutrino number density to $5.79\%$. In contrast, photon injection between neutrino decoupling and deuterium formation can access a wider range of neutrino densities~\cite{Escudero:2026mgw}. To obtain more generic constraints on the neutrino masses, we conduct a second analysis that treats the C$\nu$B temperature as a free parameter to isolate the impact of changing the neutrino abundance on the inferred neutrino-mass bounds.

Both frameworks are investigated with a Markov chain Monte Carlo~(MCMC) analysis to determine how changes to the fiducial neutrino-to-photon ratio impact cosmological limits on the neutrino masses. All models are constrained by observations of primary CMB anisotropies from \textit{Planck}~\cite{Planck:2019nip}, CMB lensing reconstructions from \textit{Planck}~\cite{Carron:2022eyg} and ACT~\cite{ACT:2023dou, ACT:2023kun}, and BAO data from DESI~\cite{DESI:2025zpo, DESI:2025zgx}. The $Y$-decay analysis is also constrained by spectral distortion limits from the Far Infrared Absolute Spectrophotometer~(FIRAS) on board the Cosmic Background Explorer~(COBE)~\cite{Fixsen:1996nj} and measurements of the primordial deuterium abundance~\cite{Cooke:2017cwo}.

We find that bounds on the neutrino masses from cosmology are robust to changes in the thermal history before recombination. The most restrictive neutrino-mass constraints arise in $Y$-decay scenarios with only photon injection, despite these decays producing the greatest reduction in the neutrino density for a given abundance of $Y$ particles. Combining all datasets, the bound tightens from $\sum m_{\nu} < \qty{0.0691}{\electronvolt}$~(95\%~CrI;~DO) for a standard thermal history to $\sum m_{\nu} < \qty{0.0652}{\electronvolt}$~(95\%~CrI;~DO) for decays into photons exclusively. Allowing the injection of both photons and DR and marginalizing over the branching ratio yields the slightly relaxed limit of $\sum m_{\nu} < \qty{0.0710}{\electronvolt}$~(95\%~CrI;~DO). Under the NO and IO, decays that inject DR in addition to photons also produce the weakest neutrino-mass constraints, although the loosening relative to $\Lambda$CDM remains marginal. When broadening the possible range of neutrino abundances by varying the temperature of the C$\nu$B directly, we obtain $\sum m_{\nu} < \qty{0.0724}{\electronvolt}$~(95\%~CrI;~DO), showing that constraints on the $Y$-decay scenario do not drive the stringency of our neutrino-mass bounds. We identify a surprising positive correlation between the neutrino abundance and neutrino masses, as a larger abundance increases the pre-recombination radiation density and mitigates the effects of massive neutrinos on the CMB.

The remainder of this paper is structured as follows. In \cref{sec:neutrino_abundance} we discuss the consequences of a reduced neutrino abundance on cosmological observables. \cref{sec:decaying_particle_model} outlines the model and cosmological effects of the decaying particle. In \cref{sec:analysis_method} we detail our analysis method, MCMC implementation, and selection of datasets. \cref{sec:results} consists of the results and discussion, and we conclude in \cref{sec:summary_and_conclusions}. The appendices provide Pearson and partial correlation coefficients for the MCMC results~(\cref{sec:correlation_coefficients}), examine prior-volume effects from different treatments of the physical mass orderings~(\cref{sec:prior_volume_effects}), and present additional MCMC results, including updated constraints on entropy-injection scenarios for fixed neutrino masses (\cref{sec:additional_MCMC_results}). Throughout this work, we use natural units ($c = \hbar = k_{\mathrm{B}} = 1$).

\section{Decreasing the neutrino abundance}
\label{sec:neutrino_abundance}

\begin{figure}
    \includegraphics[width = \columnwidth]{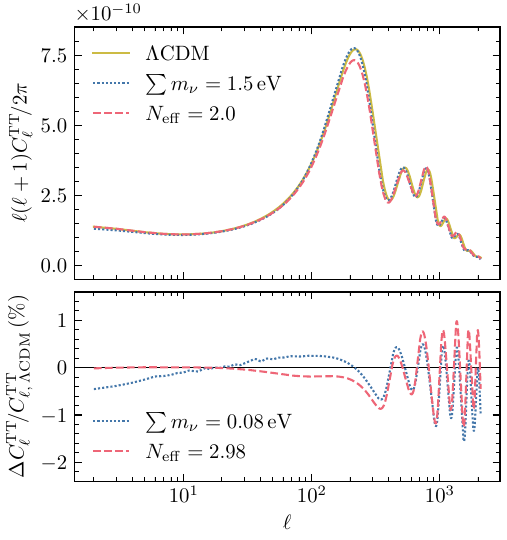}
    \caption{Effect of increasing $\sum m_{\nu}$ or reducing $N_{\mathrm{eff}}$ on the CMB TT anisotropy power spectrum, where $\Lambda$CDM corresponds to $\sum m_{\nu} = 0$ and $N_{\mathrm{eff}} = 3.044$. Dotted blue lines represent models with massive neutrinos and $N_{\mathrm{eff}} = 3.044$, while the dashed red lines correspond to cosmologies with reduced $N_{\mathrm{eff}}$ and $\sum m_{\nu} = 0$. Both modifications shift the acoustic peak locations to larger scales if all other $\Lambda$CDM parameters are held fixed. The upper panel uses exaggerated changes in $\sum m_{\nu}$ and $N_{\mathrm{eff}}$ to make deviations from the fiducial spectrum clearer, while the lower panel shows percent residuals for more realistic values, with $\Delta C_{\ell}^{\mathrm{TT}} =  C_{\ell}^{\mathrm{TT}} - C_{\ell, \, \Lambda\mathrm{CDM}}^{\mathrm{TT}}$.}
    \label{fig:spectra_comparison}
\end{figure}

Cosmology constrains $\sum m_{\nu}$ through the effects of neutrinos on the expansion rate $H$ and the growth of small-scale structure. The amplitudes of both signatures are mainly determined by the neutrino density fraction,
\begin{equation}
    f_{\nu} \equiv \frac{\rho_{\nu, 0}}{\rho_{\mathrm{m, 0}}}, 
    \label{eq:neutrino_density_fraction}
\end{equation}
where $\rho_{\nu, 0}$ and $\rho_{\mathrm{m}, 0}$ are the present-day energy densities of neutrinos and all matter, respectively~\cite{Lesgourgues:2006nd}. If $H_0$ and the other cosmological parameters are held fixed, increasing $f_{\nu}$ yields a larger value of $\Omega_{\mathrm{m}} \equiv (8 \pi G / 3 H_0^2) \rho_{\mathrm{m}, 0}$ and prolongs matter domination~(MD). Increasing $\Omega_{\mathrm{m}}$ reduces the distance to the last scattering surface $r_{\ast}$, requiring the inferred value of $h \equiv H_0 / (\qty{100}{\kilo\meter\per\second\per\mega\parsec})$ to decrease to preserve $\theta_{\mathrm{s}}$, the angular size of the sound horizon at recombination. Fixing $\theta_{\mathrm{s}}$ therefore drives a steep increase of $\Omega_{\mathrm{m}}$ with $f_{\nu}$, so that $\Omega_{\mathrm{m}} \propto (1 + f_{\nu})^5$ if the physical CDM and baryon densities, $\omega_{\mathrm{cdm}} = \Omega_{\mathrm{cdm}} h^2$ and $\omega_{\mathrm{b}} = \Omega_{\mathrm{b}} h^2$, are also held constant~\cite{Sutherland:2018ghu, Loverde:2024nfi}. The resulting delay in the onset of dark energy domination~($\Lambda$D) partially compensates for the suppression of power from neutrino free streaming, which reduces the matter power spectrum $P(k)$  on scales smaller than the neutrino free-streaming length by a factor of $1 - 8 f_{\nu}$ relative to a massless-neutrino cosmology with the same $\Omega_{\mathrm{m}}$~\cite{Hu:1997mj}. 

\begin{figure}
    \includegraphics[width = \columnwidth]{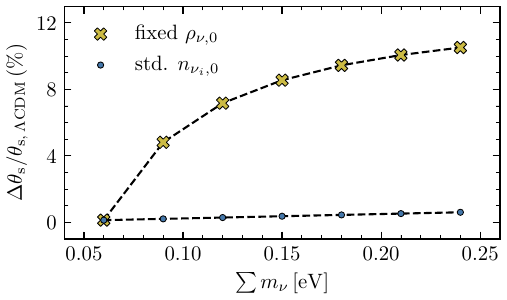}
    \caption{Percent change in $\theta_{\mathrm{s}}$ induced by increasing the sum of the neutrino masses, relative to $\Lambda$CDM with massless neutrinos. We define $\Delta \theta_{\mathrm{s}} \equiv \theta_{\mathrm{s}} - \theta_{\mathrm{s}, \, \Lambda\mathrm{CDM}}$ and fix $h = 0.6727$, $\omega_{\mathrm{cdm}} = 0.1202$, and $100 \omega_{\mathrm{b}} = 2.236$ in all models. The blue dots represent cosmologies with the standard neutrino-to-photon ratio, $n_{\nu_i} \simeq (3 / 11) n_{\gamma}$. Yellow crosses show cosmologies that fix $\rho_{\nu, 0} = \qty{6.784}{\electronvolt\per\centi\meter\cubed}$, requiring a lower C$\nu$B temperature as $\sum m_{\nu}$ increases. Though reducing the neutrino abundance might be expected to relax the bound on $\sum m_{\nu}$, the induced shift in $\theta_{\mathrm{s}}$ must be compensated by a much larger reduction in $h$ than in models with the standard $n_{\nu_i, 0}$.}
    \label{fig:theta_s_comparison}
\end{figure}

At least two neutrino mass eigenstates are now nonrelativistic, so the total neutrino density today is $n_{\nu_i, 0} \sum m_{\nu}$ if each of the three mass eigenstates has the same present-day number density, $n_{\nu_i, 0}$. If cosmological observations only constrained $f_{\nu}$, reducing the neutrino abundance would weaken the inferred bound on $\sum m_{\nu}$. Yet by lowering the radiation density before recombination, a smaller neutrino abundance also leaves an imprint on the CMB through $N_{\mathrm{eff}}$, the effective number of relativistic species evaluated at recombination. It is defined by the relation
\begin{equation}
    \rho_{\mathrm{r}} \equiv \left[1 + \frac{7}{8} \left(\frac{4}{11}\right)^{4/3} N_{\mathrm{eff}}\right] \rho_{\gamma},
    \label{eq:n_eff}
\end{equation}
where $\rho_{\mathrm{r}}$ is the total radiation energy density and $\rho_{\gamma}$ is the photon energy density. If neutrinos are thermal relics, a smaller $n_{\nu_i, 0}$ corresponds to a lower neutrino temperature and hence a reduced $N_{\mathrm{eff}}$.

\begin{figure}
    \includegraphics[width = \columnwidth]{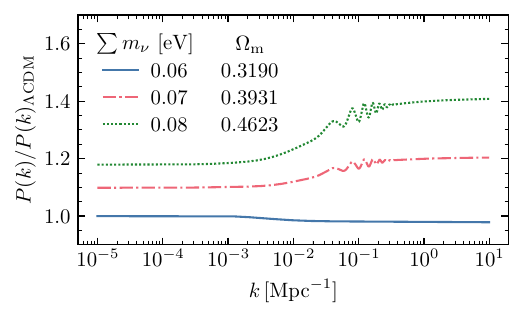}
    \caption{Relative matter power spectra for three massive-neutrino models, compared to $\Lambda$CDM with $\sum m_{\nu} = 0$ and $N_{\mathrm{eff}} = 3.044$. We fix $100 \theta_{\mathrm{s}} = 1.04090$, $\omega_{\mathrm{cdm}} = 0.1202$ and $100 \omega_{\mathrm{b}} = 2.236$ in all cosmologies, which gives $\Omega_{\mathrm{m}} = 0.3133$ in $\Lambda$CDM. We also enforce $\rho_{\nu, 0} = \qty{6.784}{\electronvolt\per\centi\meter\cubed}$ in models with massive neutrinos, matching the fiducial present-day energy density for $\sum m_{\nu} = \qty{0.06}{\electronvolt}$. The $\sum m_{\nu} = \qty{0.06}{\electronvolt}$ curve exhibits the typical massive-neutrino suppression relative to a $\Lambda$CDM cosmology with massless neutrinos and equivalent $N_{\mathrm{eff}}$. For the models with heavier neutrinos, holding $\rho_{\nu, 0}$ fixed requires a colder neutrino population and therefore a lower $N_{\mathrm{eff}}$. The reduced radiation density then shifts MR equality earlier, enhancing power on scales that are subhorizon at equality in $\Lambda$CDM. Keeping $\theta_{\mathrm{s}}$ fixed requires $h$ to decrease significantly, and the resulting increase in $\Omega_{\mathrm{m}}$ delays the onset of $\Lambda$D, which further enhances power on all scales.}
    \label{fig:mps_enhanced}
\end{figure}

A smaller value of $N_{\mathrm{eff}}$ affects the cosmological bound on $\rho_{\nu, 0}$ in two ways. As illustrated by the CMB power spectra in \cref{fig:spectra_comparison}, reducing $N_{\mathrm{eff}}$ produces the same phase shift in the acoustic peaks as increasing $\sum m_{\nu}$, since both changes increase $\theta_{\mathrm{s}}$. Decreasing $N_{\mathrm{eff}}$ lowers the early-time radiation density and reduces $H$ during radiation domination~(RD), which increases the size of the sound horizon at recombination, $r_{\mathrm{s}}$. More massive neutrinos also yield a larger $\theta_{\mathrm{s}}$, but by enhancing $H$ after recombination. When $\rho_{\nu, 0}$ is fixed, these two effects act together, exacerbating the increase in $\theta_{\mathrm{s}}$ relative to models in which the standard neutrino abundance is preserved. In \cref{fig:theta_s_comparison}, we depict the percent change in $\theta_{\mathrm{s}}$ for different values of $\sum m_{\nu}$ under these two fixed-parameter conventions. The smaller radiation density at fixed $\rho_{\nu, 0}$ leads to a significantly larger shift in $\theta_{\mathrm{s}}$ than for standard $n_{\nu_i, 0}$, as the latter leaves $N_{\mathrm{eff}}$ unchanged. Although decreasing $h$ can keep $\theta_{\mathrm{s}}$ fixed at its original value, the resulting increase in $\Omega_{\mathrm{m}}$ is strongly disfavored by DESI BAO data~\cite{Wu:2020nxz}.

\begin{figure}
    \includegraphics[width = \columnwidth]{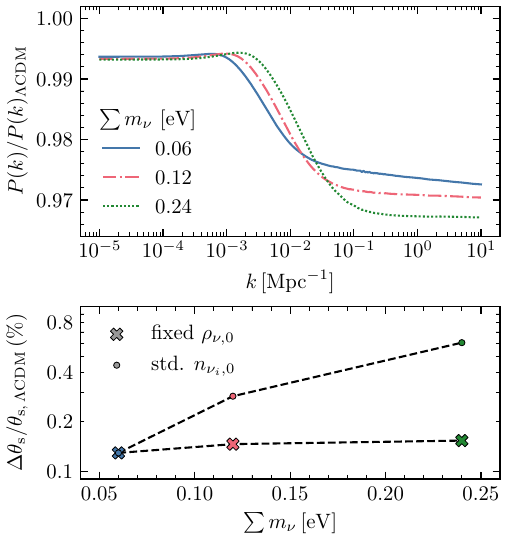}
    \caption{\textbf{Upper:} Relative matter power spectra for different massive-neutrino models, compared to $\Lambda$CDM with massless neutrinos. All cosmologies share $h = 0.6727$, $\omega_{\mathrm{cdm}} = 0.1202$, $100 \omega_{\mathrm{b}} = 2.236$, and $N_{\mathrm{eff}} = 3.044$. For models with massive neutrinos, we fix $\rho_{\nu, 0} = \qty{6.784}{\electronvolt\per\centi\meter\cubed}$ and add DR as $\sum m_{\nu}$ increases to maintain $N_{\mathrm{eff}} = 3.044$. The offset from unity on large scales reflects the delayed horizon entry relative to the massless baseline. Although all massive-neutrino cosmologies have the same $N_{\mathrm{eff}}$ and $\rho_{\nu, 0}$, larger $\sum m_{\nu}$ produces more small-scale suppression. \textbf{Lower:} Percent change in $\theta_{\mathrm{s}}$ from massive neutrinos for the same cosmological parameters, with $\Delta \theta_{\mathrm{s}} \equiv \theta_{\mathrm{s}} - \theta_{\mathrm{s}, \, \Lambda\mathrm{CDM}}$. Crosses represent models with fixed $\rho_{\nu, 0} = \qty{6.784}{\electronvolt\per\centi\meter\cubed}$, while dots show cosmologies with the standard neutrino-to-photon ratio $n_{\nu_i} \simeq (3 / 11) n_{\gamma}$. The models with standard $n_{\nu_i, 0}$ exhibit larger deviations in $\theta_{\mathrm{s}}$ because $\rho_{\mathrm{m}, 0}$ increases with $\sum m_{\nu}$, but the phase shift still increases with heavier neutrinos even if $N_{\mathrm{eff}}$ and $\rho_{\nu, 0}$ are fixed.}
    \label{fig:mps_theta_s_comparison}
\end{figure}

Reducing $N_{\mathrm{eff}}$ also impacts $P(k)$. In \cref{fig:mps_enhanced}, we compare massive-neutrino models against $\Lambda$CDM with massless neutrinos, while holding $\theta_{\mathrm{s}}$, $\omega_{\mathrm{cdm}}$, and $\omega_{\mathrm{b}}$ constant. We also fix $\rho_{\nu, 0}  = \qty{6.784}{\electronvolt\per\centi\meter\cubed}$ when including massive neutrinos, which corresponds to the fiducial value for $\sum m_{\nu} = \qty{0.06}{\electronvolt}$ and ensures that all cosmologies have $f_{\nu} = \num{4.519e-3}$. The $\sum m_{\nu} = \qty{0.06}{\electronvolt}$ curve exhibits the typical small-scale suppression from massive neutrinos, as $N_{\mathrm{eff}}$ in this model is equal to that in $\Lambda$CDM. The cosmologies with larger $\sum m_{\nu}$ have reduced $N_{\mathrm{eff}}$, which shifts MR equality to earlier times and enhances power on scales that are subhorizon at MR equality in $\Lambda$CDM. At fixed $\theta_{\mathrm{s}}$, the decrease in $h$ for the heavier-neutrino models delays the onset of $\Lambda$D, which enhances $P(k)$ across all scales. Although primary CMB observations from \textit{Planck} favor moderately enhanced lensing power~\cite{Addison:2023fqc}, \cref{fig:mps_enhanced} shows how increasing $\sum m_{\nu}$ from $\qty{0.06}{\electronvolt}$ to $\qty{0.08}{\electronvolt}$ at fixed $\theta_{\mathrm{s}}$ and $\rho_{\nu, 0}$ excessively enhances $P(k)$.

The addition of DR would permit a smaller neutrino abundance without reducing $N_{\mathrm{eff}}$. However, even if $N_{\mathrm{eff}}$ and $\rho_{\nu, 0}$ are held fixed, varying the sum of the neutrino masses still affects $P(k)$ and the CMB. The upper panel of \cref{fig:mps_theta_s_comparison} compares $P(k)$ for various massive-neutrino cosmologies with the same $\rho_{\nu, 0}$ to $P(k)$ with massless neutrinos, though all cosmologies have the same $N_{\mathrm{eff}}$. Since all massive-neutrino models share $f_{\nu} = \num{4.519e-3}$, differences in the depth and characteristic scale of the step-like suppression arise from the mass dependence of the neutrino nonrelativistic transition~\cite{Agarwal:2010mt}. For a single mass eigenstate, this occurs at a redshift $z$ given by
\begin{equation}
    1 + z_{\mathrm{nr}} \simeq 151 \left(\frac{m_{\nu}}{\qty{0.08}{\electronvolt}}\right) \left(\frac{0.71611}{T_{\nu} / T_{\gamma}}\right).
    \label{eq:z_nonrelativistic}
\end{equation}
Here, $T_{\nu}$ and $T_{\gamma}$ denote the neutrino and photon temperatures, with their ratio in \cref{eq:z_nonrelativistic} evaluated while the neutrinos are still relativistic. More massive neutrinos become nonrelativistic earlier and therefore contribute to the matter density over a longer span of cosmic history. The characteristic wavenumber above which neutrinos suppress $P(k)$ also depends on their mass, as it is set by the size of the comoving horizon when each massive eigenstate becomes nonrelativistic, which yields
\begin{equation}
    k_{\mathrm{nr}} \simeq \num{3.34e-4} \sqrt{\Omega_{\mathrm{m}} h^2 (1 + z_{\mathrm{nr}})} \, \unit{\per\mega\parsec}.
    \label{eq:k_nonrelativistic}
\end{equation}
Accounting for these effects, the suppression in $P(k)$ for fixed $\Omega_{\mathrm{m}}$ is then given by
\begin{equation}
    \frac{P(k)}{P(k)_{\Lambda\mathrm{CDM}}} \simeq 1 - 2 f_{\nu} - \frac{6}{5} f_{\nu} \ln(1 + z_{\mathrm{nr}}),
    \label{eq:mps_suppression}
\end{equation}
which makes the dependence on $z_{\mathrm{nr}}$ explicit~\cite{Green:2024xbb}. The $-2 f_{\nu}$ term reflects the fraction of matter that does not cluster for $k > k_{\mathrm{nr}}$, while the logarithmic term encodes the suppressed growth of CDM and baryon perturbations due to the presence of massive neutrinos.\footnote{Under the assumption of a standard thermal history, relic neutrinos with a mass of $m_{\nu} \sim \qty{0.08}{\electronvolt}$ become nonrelativistic at $z_{\mathrm{nr}} \sim 150$, for which \cref{eq:mps_suppression} reduces to the familiar leading-order approximation in $f_{\nu}$, $P(k) / P(k)_{\Lambda\mathrm{CDM}} \approx 1 - 8 f_{\nu}$.}

Since increasing $m_{\nu}$ shifts the neutrino nonrelativistic transition to higher redshift, $\theta_{\mathrm{s}}$ is also affected. For the models with fixed $\rho_{\nu, 0}$ in \cref{fig:mps_theta_s_comparison}, the changing value of $\theta_{\mathrm{s}}$ in the lower panel reflects the different redshifts at which the neutrinos become nonrelativistic. After this transition occurs, the energy density of a massive neutrino species redshifts as matter rather than radiation, so that the total energy density of the universe becomes larger relative to that of a lighter-mass model whose neutrinos remain relativistic for longer. This discrepancy enhances $H$ after recombination for larger $\sum m_{\nu}$, but as these models differ only over the interval spanning their respective values of $z_{\mathrm{nr}}$, the resulting increase in $\theta_{\mathrm{s}}$ from heavier neutrinos remains subpercent at fixed $N_{\mathrm{eff}}$. The change in $\theta_{\mathrm{s}}$ is significantly smaller than in the standard-abundance case, where larger values of $\sum m_{\nu}$ also increase $\rho_{\mathrm{m}, 0}$ relative to $\Lambda$CDM, more substantially enhancing the post-recombination expansion rate.

These various results predict that decreasing $n_{\nu_i, 0}$ will tighten the bound on $\rho_{\nu, 0}$, since reducing $N_{\mathrm{eff}}$ exacerbates the impact of massive neutrinos on the CMB. When the decrease in the relativistic neutrino density is compensated by DR, massive neutrinos still suppress CMB lensing relative to $\Lambda$CDM and produce a small, residual shift in $\theta_{\mathrm{s}}$. Even at fixed $\rho_{\nu, 0}$, the magnitude of both effects depends on the mass of each massive neutrino eigenstate, which 
determines the timing of its corresponding nonrelativistic transition. Cosmological observables therefore depend on a nontrivial interplay between the neutrino abundance, $N_{\mathrm{eff}}$, and the individual neutrino masses, which together prevent a reduced neutrino abundance from relaxing the upper bound on $\sum m_{\nu}$.

\section{Decaying particle model}
\label{sec:decaying_particle_model}

To explore the impact of reducing the neutrino abundance within a physically viable framework, we adopt a three-fluid model in which a subdominant hidden-sector $Y$ particle decays into photons and DR. A branching ratio determines the relative energy deposited into photons and controls whether the net change in $N_{\mathrm{eff}}$ from its fiducial value is positive or negative. Photon injection after neutrino decoupling lowers the neutrino abundance relative to the photon bath and thereby reduces $N_{\mathrm{eff}}$, while the DR increases $N_{\mathrm{eff}}$ without altering the neutrino number density. Implementing these changes through a physical model with photon injection introduces additional constraints from limits on CMB spectral distortions and measurements of the primordial deuterium abundance.

\subsection{Background evolution}
\label{subsec:background}

The background equations for the energy densities of the $Y$ particles ($\rho_Y$), photons ($\rho_{\gamma}$), and DR ($\rho_{\mathrm{dr}}$) are
\begin{subequations}
\label{eq:background_equations}
\begin{align}
    \dot{\rho}_Y + 3 \frac{\dot{a}}{a} \rho_Y &= -a \Gamma_Y \rho_Y,
    \label{eq:y_evolution}\\
    \dot{\rho}_{\gamma} + 4 \frac{\dot{a}}{a} \rho_{\gamma} &= f_{\gamma} a \Gamma_Y \rho_Y,
    \label{eq:photon_evolution}\\
    \dot{\rho}_{\mathrm{dr}} + 4 \frac{\dot{a}}{a} \rho_{\mathrm{dr}} &= (1 - f_{\gamma}) a \Gamma_Y \rho_Y.
    \label{eq:dr_evolution}
\end{align}
\end{subequations}
An overdot denotes a derivative with respect to conformal time $\tau$, $a$ is the scale factor, $\Gamma_Y$ is the decay rate, and $f_{\gamma}$ is the fraction of $Y$ particles that decay into photons. We parameterize $\Gamma_Y$ by a reheating temperature $T_{\mathrm{RH}}$, defined by evaluating the Friedmann equation for RD at $H = \Gamma_Y$, where $H(a) \equiv \dot{a} / a^2$ is the standard Hubble parameter. This treatment yields
\begin{equation}
    \Gamma_Y \equiv \sqrt{\frac{8 \pi}{3 m_{\mathrm{Pl}}^2} \left(\frac{\pi^2}{30} g_{\ast} T_{\mathrm{RH}}^4\right)},
    \label{eq:decay_rate}
\end{equation}
where $m_{\mathrm{Pl}}$ is the Planck mass and $g_{\ast} = 3.38$ is the effective number of relativistic degrees of freedom after BBN. 

\begin{figure}
    \includegraphics[width = \columnwidth]{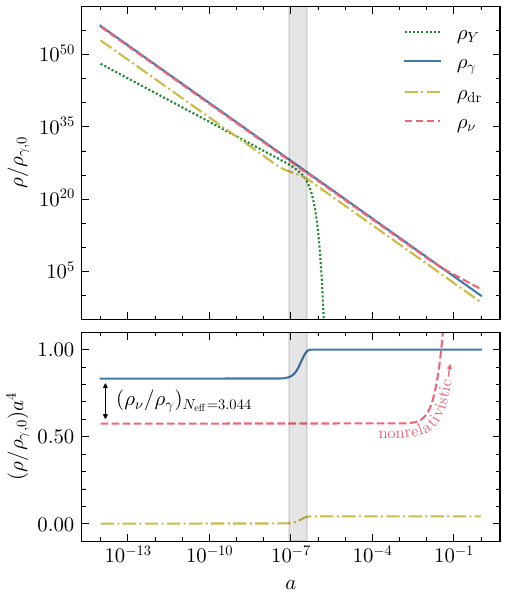 }
    \caption{Background evolution of the energy densities of the $Y$ particles, its decay products, and massive neutrinos. Each is normalized to the present-day photon energy density, $\rho_{\gamma, 0}$. We use representative decay parameters of $f_{\gamma} = 0.8$, $\Gamma_Y = \qty{e-6}{\per\second}$, and $R_{\Gamma} = 0.07$, and denote the decay epoch with the gray vertical band. Initially, $\rho_Y$ scales like matter until most of the $Y$ particles decay at $a_{\mathrm{RH}} \approx \num{1.4e-7}$, at which time $\rho_Y$ rapidly decreases. Both $\rho_{\gamma}$ and $\rho_{\mathrm{dr}}$ evolve as standard radiation except during the decay, while $\rho_{\nu}$ scales as radiation until late times when the neutrinos become nonrelativistic. We emphasize this behavior in the lower panel by plotting the normalized comoving energy densities. Photon injection reduces the neutrino-to-photon ratio from its value for $N_{\mathrm{eff}} = 3.044$, with the widening gap between the solid blue and dashed red curves indicating fewer neutrinos per photon.}
    \label{fig:background_evolution}
\end{figure}

During RD, the expansion rate is approximated by $H(a) \approx H_{\mathrm{i}} (a_{\mathrm{i}} / a)^{2}$, where $H_{\mathrm{i}} \equiv H(a_{\mathrm{i}})$. We can then solve \cref{eq:y_evolution} analytically to determine $\rho_Y(a)$,
\begin{equation}
    \rho_Y(a) = \rho_{Y, {\mathrm{i}}} \left(\frac{a_{\mathrm{i}}}{a}\right)^3 \mathrm{e}^{\tilde{\Gamma}_Y [1 - (a / a_{\mathrm{i}})^2] / 2},
    \label{eq:y_energy_density}
\end{equation}
where $\rho_{Y, \mathrm{i}} \equiv \rho_Y(a_{\mathrm{i}})$ and $\tilde{\Gamma}_Y \equiv \Gamma_Y / H_{\mathrm{i}}$. The scale factor characterizing the decay epoch is defined by $\smash{a_{\mathrm{RH}} / a_{\mathrm{i}} \equiv \tilde{\Gamma}_Y^{-1 / 2}}$, which implies $H(a_{\mathrm{RH}}) \simeq \Gamma_Y$ if $\rho_Y \ll \rho_{\mathrm{r}}$. Neglecting the energy injection into radiation due to $Y$ decay gives $\rho_{\mathrm{r}}(a_{\mathrm{RH}}) \approx \rho_{\mathrm{r, i}} (a_{\mathrm{i}} / a_{\mathrm{RH}})^4$. Under these assumptions, it follows from \cref{eq:y_energy_density} that the maximum value of $\rho_Y / \rho_{\mathrm{r}}$ occurs at $a_{\mathrm{RH}}$. We therefore define
\begin{equation}
    R_{\Gamma} \equiv \frac{\rho_{Y, \mathrm{i}}}{\rho_{\mathrm{r, i}}} \sqrt{\frac{\mathrm{e}^{\tilde{\Gamma}_Y - 1}}{\tilde{\Gamma}_Y}} \simeq \frac{\rho_Y(a_{\mathrm{RH}})}{\rho_{\mathrm{r}}(a_{\mathrm{RH})}},
    \label{eq:max_ratio}
\end{equation}
and use $R_{\Gamma}$ to describe the initial abundance of $Y$ particles. Since $Y$ decay occurs while the neutrinos are still relativistic, the initial radiation energy density is given by $\rho_{\mathrm{r, i}} = \rho_{\gamma, \mathrm{i}} + \rho_{\mathrm{dr, i}} + \rho_{\nu, \mathrm{i}}$. In \cref{fig:background_evolution}, we depict the background evolution of the energy densities for each component in the three-fluid model and for massive neutrinos within a representative decay scenario. Entropy injection during $Y$ decay modifies the adiabatic evolution of the photon and DR energy densities near $a_{\mathrm{RH}}$, at which point the $Y$ particles decay away and their energy density rapidly decreases. The energy density of the neutrinos initially evolves as radiation, but subsequently scales as matter after the neutrinos become nonrelativistic.

The $Y$ decay injects no neutrinos, and we model the relic neutrino background as three active species with Fermi--Dirac distributions. In reality, neutrinos were partially heated by electron--positron annihilation because their decoupling was not instantaneous, producing small distortions in their thermal spectra. A single neutrino temperature $T_{\nu}$ therefore cannot reproduce the correct relativistic energy density and number density simultaneously. The nonthermal corrections to the relativistic neutrino energy density are typically incorporated by adopting $N_{\mathrm{eff}}(\mathrm{BBN}) = 3.044$~\cite{Froustey:2020mcq, Bennett:2020zkv}. Once nonrelativistic, the neutrino energy density is dominated by their rest mass, and the fiducial neutrino abundance fixes the late-time relation $\sum m_{\nu} / \omega_{\nu} = \qty{93.14}{\electronvolt}$~\cite{Lesgourgues:2012uu}, where $\omega_{\nu}$ is the neutrino physical density parameter. We enforce an initial neutrino temperature of $T_{\nu, \mathrm{i}} = 0.71611 T_{\gamma, \mathrm{i}}$, which reproduces the correct late-time number density of relic neutrinos within the Fermi--Dirac approximation. To parameterize the resulting radiation content, we define the effective number of species for either neutrinos or DR as
\begin{equation}
    N_{\nu, \, \mathrm{dr}} \equiv \frac{8}{7} \left(\frac{11}{4}\right)^{4/3} \frac{\rho_{\nu, \, \mathrm{dr}}}{\rho_{\gamma}}.
    \label{n_neutrino_dr}
\end{equation}
The initial value of $T_{\nu}$ that we impose corresponds to $N_{\nu}(\mathrm{BBN}) = 3.0396$. To reproduce $N_{\mathrm{eff}}(\mathrm{BBN}) = 3.044$, we include a primordial component of DR parameterized by $N_{\mathrm{dr}}(\mathrm{BBN}) = 0.00441$. Any additional DR is sourced only by injection from the decaying $Y$ particles.

Since observations of the CMB fix the present-day photon temperature, the photon bath must be initialized below its fiducial temperature so that the post-decay evolution recovers the measured value of $T_{\gamma, 0} = \qty{2.7255}{\kelvin}$. This nonstandard behavior is evident in the lower panel of \cref{fig:background_evolution}, which depicts the comoving energy densities of all relativistic species normalized to the fiducial photon energy density today. The curve depicting the comoving photon energy density begins below unity and increases to the standard value for $T_{\gamma, 0} = \qty{2.7255}{\kelvin}$ after the entropy injection from $Y$ decay occurs. As in Ref.~\cite{Sobotka:2022vrr}, we parameterize the required adjustment to the initial photon temperature by defining the gap ratio $g \equiv \rho_{\mathrm{r, f}} a_{\mathrm{f}}^4 / \rho_{\mathrm{r, i}} a_{\mathrm{i}}^4$, where $\rho_{\mathrm{r, f}} a_{\mathrm{f}}^4$ and $\rho_{\mathrm{r, i}} a_{\mathrm{i}}^4$ denote the comoving energy densities of all radiation after and before $Y$ decay, respectively. As $g$ only depends on the energy density of the $Y$ particles when $H \sim \Gamma_Y$, it is independent of the reheating temperature and a simple function of $R_{\Gamma}$, which we fit with a fourth-order polynomial,
\begin{equation}
    g \simeq 0.0274 R_{\Gamma}^4 - 0.152 R_{\Gamma}^3 + 0.646 R_{\Gamma}^2 + 2.07 R_{\Gamma} + 1.
    \label{eq:gap_ratio}
\end{equation}
We also define the gap ratios for each decay product, $g_{\gamma} \equiv \rho_{\gamma, \mathrm{f}} a_{\mathrm{f}}^4 / \rho_{\gamma, \mathrm{i}} a_\mathrm{i}^4$ and $g_{\mathrm{dr}} \equiv \rho_{\mathrm{dr, f}} a_{\mathrm{f}}^4 / \rho_{\mathrm{dr, i}} a_{\mathrm{i}}^4$. In terms of the decay parameters, these are given by
\begin{subequations}
\label{eq:decay_product_gap_ratios}
\begin{align}
    g_{\gamma} &= f_{\gamma} (g - 1) \left[1 + \frac{7}{8} (3.044) \left(\frac{4}{11}\right)^{4/3}\right] + 1,
    \label{eq:photon_gap_ratio}\\
    g_{\mathrm{dr}} &= \frac{(1 - f_{\gamma}) (g - 1)}{0.00441} \left[3.044 + \frac{8}{7} \left(\frac{11}{4}\right)^{4/3}\right] + 1.
    \label{eq:dr_gap_ratio}
\end{align}
\end{subequations}

If $a_0$ is the scale factor today, the modification to the initial photon temperature is
\begin{equation}
    T_{\gamma, \mathrm{i}} a_{\mathrm{i}} = \frac{1}{g_{\gamma}^{1/4}} T_{\gamma, 0} a_0,
    \label{eq:initial_photon_temperature}
\end{equation}
which is completely determined by $R_{\Gamma}$ and $f_{\gamma}$ by using \cref{eq:gap_ratio,eq:photon_gap_ratio}. Since we set $T_{\nu, \mathrm{i}} = 0.71611 T_{\gamma, \mathrm{i}}$, lowering the initial photon temperature yields a smaller relic neutrino abundance than for a standard thermal history. Entropy injection heats the photon bath without affecting neutrinos, so that after the decay, the neutrino temperature relative to photons is given by
\begin{equation}
    T_{\nu, \mathrm{f}} = \frac{0.71611}{g_{\gamma}^{1/4}} T_{\gamma, \mathrm{f}}.
    \label{eq:final_neutrino_temperature}
\end{equation}

Having specified $\rho_{\mathrm{r, i}}$, $H_{\mathrm{i}}$ is also determined, which allows $\rho_{Y, \mathrm{i}}$ to be calculated from \cref{eq:max_ratio}. The initial conditions for our three-fluid model are therefore completely constrained by the three decay parameters, $\{f_{\gamma}, \Gamma_Y, R_{\Gamma}\}$, which are sampled in our MCMC analysis. We modify the Boltzmann solver \texttt{CLASS}~\cite{Blas:2011rf} by adding the $Y$ particles as a new species and by reducing the standard initial values of $T_{\gamma}$ and $T_{\nu}$. The \texttt{CLASS} module for non-cold relics~\cite{Lesgourgues:2011rh} is used to implement massive neutrinos. 

\subsection{Perturbations}
\label{subsec:perturbations}

We obtain evolution equations for the perturbations in our three fluids by perturbing covariant versions of \cref{eq:background_equations}. Details of the derivation can be found in Refs.~\cite{Erickcek:2011us, Sobotka:2023bzr}. We work in the conformal Newtonian gauge, in which the perturbed Friedmann--Lema\^{\i}tre--Robertson--Walker~(FLRW) metric is given by
\begin{equation}
    \mathrm{d}s^2 = a^2(\tau) [-(1 + 2 \psi) \mathrm{d}\tau^2 + \delta_{ij} (1 - 2 \phi) \mathrm{d}x^i \mathrm{d}x^j],
    \label{eq:conformal_newtonian_flrw_metric}
\end{equation}
where $\psi$ and $\phi$ are the metric scalar perturbations. For each species, we also define
\begin{subequations}
\label{eq:perturbation_variables}
\begin{align}
    \delta_n &\equiv \frac{\delta \rho_n(\tau, \vec{x})}{\bar{\rho}_n(\tau)},
    \label{eq:delta_perturbation}\\
    \theta_n &\equiv \mathrm{i} k_j (v^j)_n,
    \label{eq:theta_perturbation}
\end{align}
\end{subequations}
where $\delta \rho \equiv \rho - \bar{\rho}$ is the density perturbation, $\bar{\rho}$ is the homogeneous background density, and $v^j \equiv \mathrm{d}x^j / \mathrm{d}\tau$ is the peculiar fluid velocity. The perturbation equations for the three fluids in Fourier space are then
\begin{subequations}
\label{eq:perturbation_equations}
\begin{widetext}
\begin{align}
    \dot{\delta}_Y + \theta_Y - 3 \dot{\phi} &= -a \Gamma_Y \psi,
    \label{eq:delta_y_evolution}\\
    \dot{\theta}_Y + \mathcal{H} \theta_Y - k^2 \psi &= 0,
    \label{eq:theta_y_evolution}\\
    \dot{\delta}_{\gamma} + \frac{4}{3} \theta_{\gamma} - 4 \dot{\phi} &= f_{\gamma} a \Gamma_Y \frac{\bar{\rho}_Y}{\bar{\rho}_{\gamma}} (\delta_Y - \delta_{\gamma} + \psi),
    \label{eq:delta_photon_evolution}\\
    \dot{\theta}_{\gamma} - k^2 \left(\frac{1}{4} \delta_{\gamma} - \sigma_{\gamma}\right) - k^2 \psi &= f_{\gamma} a \Gamma_Y \frac{\bar{\rho}_Y}{\bar{\rho}_{\gamma}} \left(\frac{3}{4} \theta_Y - \theta_{\gamma}\right) + n_{\mathrm{e}} \sigma_{\mathrm{T}} (\theta_{\mathrm{b}} - \theta_{\gamma}),
    \label{eq:theta_photon_evolution}\\
    \dot{\delta}_{\mathrm{dr}} + \frac{4}{3} \theta_{\mathrm{dr}} - 4 \dot{\phi} &= (1 - f_{\gamma}) a \Gamma_Y \frac{\bar{\rho}_Y}{\bar{\rho}_{\mathrm{dr}}} (\delta_Y - \delta_{\mathrm{dr}} + \psi),
    \label{eq:delta_dr_evolution}\\
    \dot{\theta}_{\mathrm{dr}} - k^2 \left(\frac{1}{4} \delta_{\mathrm{dr}} - \sigma_{\mathrm{dr}}\right) - k^2 \psi &= (1 - f_{\gamma}) a \Gamma_Y \frac{\bar{\rho}_Y}{\bar{\rho}_{\mathrm{dr}}} \left(\frac{3}{4} \theta_Y - \theta_{\mathrm{dr}}\right),
    \label{eq:theta_dr_evolution}
\end{align}
\end{widetext}
\end{subequations}
where $n_{\mathrm{e}}$ is the mean electron density, $\sigma_{\mathrm{T}}$ is the Thomson cross section, and $\theta_{\mathrm{b}}$ is the baryon velocity divergence. As the $Y$ particles do not couple directly to baryons, no decay source term appears in the $\theta_{\mathrm{b}}$ equation. The quantities $\sigma_{\gamma}$ and $\sigma_{\mathrm{dr}}$ denote the photon and DR shear, respectively. Their evolution equations and all higher moments of the photon and DR distributions retain their standard form. Lastly, the Einstein equations~\cite{Ma:1995ey} are
\begin{subequations}
\label{eq:einstein_equations}
\begin{align}
    k^2 \phi + 3 \frac{\dot{a}}{a} \left(\dot{\phi} + \frac{\dot{a}}{a} \psi\right) &= -4 \pi G a^2 \sum_n \bar{\rho}_n \delta_n,
    \label{eq:time_time_einstein_equation}\\
    k^2 (\phi - \psi) &= 16 \pi G a^2 \sum_n \bar{\rho}_n \sigma_n.
    \label{eq:space_space_einstein_equation}
\end{align}
\end{subequations}

The initial scale factor in \texttt{CLASS} is chosen to be well before the onset of $Y$ decay and early enough that all modes of interest are outside the horizon, implying $\tilde{\Gamma}_Y \ll 1$ and $k \ll a_{\mathrm{i}} H_{\mathrm{i}}$. All three fluids can therefore be initialized with the standard adiabatic superhorizon conditions, with the $Y$ particles assigned the initial condition of a generic decaying dark matter component~\cite{Audren:2014bca}.\footnote{Since our analysis includes an initial DR component, we impose the standard adiabatic superhorizon condition, $\delta_{\mathrm{dr}}(a_{\mathrm{i}}) = \delta_{\gamma}(a_{\mathrm{i}})$. If the DR were exclusively sourced by the decay of the $Y$ particles, a different (and gauge-dependent) relation would apply~\cite{Sobotka:2023bzr}.} We modify \texttt{CLASS} by adding the perturbation equations for the $Y$ particles and the decay source terms in the photon and DR sectors. \texttt{CLASS} contains two inequality conditions that govern the initialization of perturbation modes. Each $k$ mode begins to be integrated as soon as one of these conditions is satisfied, with the timing shifted earlier or later by varying the numerical threshold in each inequality. To implement the $Y$-decay model, we add an additional condition,
\begin{equation}
    \frac{\Gamma_Y}{H} \geq \texttt{start\_large\_k\_at\_gamma\_y\_over\_h}.
    \label{eq:k_initialization_inequality_condition}
\end{equation}
Choosing $\texttt{start\_large\_k\_at\_gamma\_y\_over\_h} \ll 1$ ensures that the perturbations begin evolving well before the effects of $Y$ decay become significant.

\begin{table}
    \caption{Numerical thresholds for the \texttt{CLASS} initialization and TCA-off conditions, chosen to ensure that all perturbation modes are initialized well before $Y$ decay becomes significant, and that the TCA is disabled immediately after the integration of each perturbation mode begins.}
    \label{tab:tca_inequality_condition_thresholds}
    \begin{ruledtabular}
    \begin{tabular}{lS[table-format = 1.1e-2]}
        \texttt{start\_small\_k\_at\_tau\_c\_over\_tau\_h}
        & 1.0e-9\\
        \texttt{start\_large\_k\_at\_tau\_h\_over\_tau\_k}
        & 5.0e-2\\
        \texttt{start\_large\_k\_at\_gamma\_y\_over\_h}
        & 1.0e-4\\
        \texttt{tight\_coupling\_trigger\_tau\_c\_over\_tau\_h}
        & 1.1e-9\\
        \texttt{tight\_coupling\_trigger\_tau\_c\_over\_tau\_k}
        & 1.0e-10\\
        \texttt{tight\_coupling\_trigger\_gamma\_y\_over\_h}
        & 1.1e-4
    \end{tabular}
    \end{ruledtabular}
\end{table}

Numerical solutions computed by \texttt{CLASS} include the full set of cosmological species. Although all direct interactions of the $Y$ particles are included in \cref{eq:perturbation_equations,eq:einstein_equations}, the photon perturbation equations take a different form in the tight-coupling approximation~(TCA), which applies when photons and baryons are tightly coupled by Thomson scattering in the early universe~\cite{Peebles:1970ag}. We found that adding the $Y$--$\gamma$ interactions to the TCA equations led to numerical instabilities on superhorizon scales, so we integrate the full perturbation system directly.\footnote{With the stiff integrators used by \texttt{CLASS}, the resulting increase in computation time is minimal~\cite{Blas:2011rf}.} \texttt{CLASS} nevertheless requires each perturbation mode to begin evolving while the TCA is active. A separate set of inequality conditions governs when the TCA is switched off for each $k$ mode, which occurs as soon as the first of these conditions is satisfied. To accommodate our modifications to the perturbation initialization scheme, we introduce an additional TCA-off condition,
\begin{equation}
    \frac{\Gamma_Y}{H} \geq \texttt{tight\_coupling\_trigger\_gamma\_y\_over\_h}.
    \label{eq:tca_off_inequality_condition}
\end{equation}
In conjunction with the standard TCA-off conditions, choosing this threshold to be slightly larger than $\texttt{start\_large\_k\_at\_gamma\_y\_over\_h}$ ensures that TCA is active when each mode begins evolving and is switched off immediately afterward. For the earliest decays and for all modes $k < \qty{100}{\hubble\per\mega\parsec}$, the thresholds listed in \cref{tab:tca_inequality_condition_thresholds} ensure that the TCA is effectively never used during integration. The standard initialization and TCA-off conditions in \texttt{CLASS} are defined in Ref.~\cite{Blas:2011rf}.

\subsection{Effects}
\label{subsec:effects}

A detailed discussion of the effects of $Y$ decay on BBN and the CMB is given in Ref.~\cite{Sobotka:2022vrr}, and we summarize the key results here. The primary impact during BBN is to modify the baryon-to-photon ratio, $\eta_{\mathrm{BBN}}$, as any $f_{\gamma} \neq 0$ decay reduces $\rho_{\gamma}$ at BBN if $T_{\gamma, 0}$ is fixed. CMB observations independently constrain the baryon-to-photon ratio at recombination, $\eta_{\ast}$. Baryon number is conserved, so the BBN ratio is larger than the late-time value by
\begin{equation}
    \eta_{\mathrm{BBN}} = g_{\gamma}^{3/4} \eta_{\ast}.
    \label{eq:eta_bbn}
\end{equation}
The energy density of the $Y$ particles also increases the expansion rate during BBN. We account for this effect and the change to $\eta_{\mathrm{BBN}}$ by modifying \texttt{PArthENoPE v3.0}~\cite{Gariazzo:2021iiu} to compute the altered abundances of helium ($Y_{\mathrm{He}}$) and deuterium (D/H). The resulting value of $Y_{\mathrm{He}}$ is passed to \texttt{CLASS} for its recombination calculations.

After photon thermalization becomes inefficient, energy injected into the photon bath creates deviations from a blackbody CMB spectrum known as spectral distortions. Injection at $\num{5e4} \lesssim z \lesssim \num{2e6}$ primarily generates $\mu$-type distortions, while injection at lower redshifts produces $y$-type distortions. At earlier times, the injected energy is efficiently thermalized into a shift of the blackbody temperature~\cite{Chluba:2011hw}. We compute the distortion parameters in \texttt{CLASS} using a spectral distortion module from Ref.~\cite{Lucca:2019rxf}. The fractional energy injection is
\begin{equation}
    \tilde{d} \equiv \left.\frac{\Delta \rho_{\gamma}}{\rho_{\gamma}}\right|_d = \int \frac{\mathrm{d}Q / \mathrm{d}z}{\rho_{\gamma}} \mathcal{J}_d(z) \mathrm{d}z,
    \label{eq:sd_fractional_energy_injection}
\end{equation}
where $\mathcal{J}_d(z)$ is the branching function that weights the contribution of energy injection at redshift $z$ to distortion type $d$. Expressing the energy injection rate $\mathrm{d}Q / \mathrm{d}z$ in terms of the $Y$-particle contribution gives
\begin{equation}
    \frac{\mathrm{d}Q / \mathrm{d}z}{\rho_{\gamma}} = -\frac{f_{\gamma} \Gamma_Y \rho_{Y}}{(1 + z) H \rho_{\gamma}}.
    \label{eq:y_energy_injection_rate}
\end{equation}
The fractional energy injections are related to the physical distortion amplitudes by $\mu = 1.4 \tilde{\mu}$ and $y = \tilde{y} / 4$. Decays considered in this work occur over the approximate redshift range $\num{e5} \lesssim z_{\mathrm{RH}} \lesssim \num{4.5e7}$, during which the bound on $\mu$-type distortions is most constraining. Ref.~\cite{Sobotka:2022vrr} found that the COBE/FIRAS bound on $\mu$ strongly restricts the timing of photon injection, requiring $f_{\gamma} \rightarrow 0$ for $T_{\mathrm{RH}} \lesssim \qty{9.5e-4}{\mega\electronvolt}$.

The injection of photons and DR changes $N_{\mathrm{eff}}$ at the time of recombination. Although the decaying $Y$ particles do not create any neutrinos, photon injection causes $\rho_{\nu} / \rho_{\gamma}$ to decrease from its fiducial value, which we enforce at BBN. The decay can also raise the DR density, so $\rho_{\mathrm{r}} / \rho_{\gamma}$ will either increase or decrease from its initial value depending on $f_{\gamma}$. After the decay, the effective number of neutrino species and that of DR species are then
\begin{subequations}
\label{eq:n_components_post_decay}
\begin{align}
    N_{\nu} &= \frac{1}{g_{\gamma}} N_{\nu}(\mathrm{BBN}),
    \label{eq:n_nu_post}\\
    N_{\mathrm{dr}} &= \frac{g_{\mathrm{dr}}}{g_{\gamma}} N_{\mathrm{dr}}(\mathrm{BBN}).
    \label{eq:n_dr_post_decay}
\end{align}
\end{subequations}
Combining these results gives $N_{\mathrm{eff}}$ after $Y$ decay and at recombination. In terms of the gap ratios, which themselves are functions of the decay parameters, we have
\begin{equation}
    N_{\mathrm{eff}} = \frac{1}{g_{\gamma}} (3.0396 + 0.00441 g_{\mathrm{dr}}).
    \label{eq:n_eff_post_decay}
\end{equation}
If $f_{\gamma} = 0.5913$, $N_{\mathrm{eff}} = 3.044$ is maintained for any $R_{\Gamma}$. Any decay scenario with $f_{\gamma} < 0.5913$ gives $N_{\mathrm{eff}} > 3.044$, while a $f_{\gamma} > 0.5913$ decay scenario yields $N_{\mathrm{eff}} < 3.044$.

\section{Analysis method}
\label{sec:analysis_method}

\subsection{Datasets and likelihoods}
\label{subsec:datasets_and_likelihoods}

Our primary CMB dataset is denoted by \textbf{\textit{Planck}} and consists of \textit{Planck} PR3 data, including the \texttt{Plik} TT,TE,EE likelihood at $\ell \geq 30$, the low-$\ell$ temperature \texttt{Commander} likelihood, and the low-$\ell$ \texttt{SimAll} EE likelihood~\cite{Planck:2019nip}. We then add the DESI DR2 BAO sample~\cite{DESI:2025zpo, DESI:2025zgx}, denoted by \textbf{DESI} and implemented with the likelihood\footnote{\url{https://github.com/LauraHerold/MontePython_desilike}} from Ref.~\cite{Herold:2025hkb}. The final dataset we include is the combined CMB lensing reconstruction likelihood from ACT DR6~\cite{ACT:2023dou, ACT:2023kun} and \textit{Planck} PR4 \texttt{NPIPE}~\cite{Carron:2022eyg}, denoted by \textbf{lens(PR4+ACT)}. We omit primary CMB data from ACT due to the discrepancy with DESI at the $3 \sigma$ level in $\Lambda$CDM. Bounds on $\sum m_{\nu}$ are also sensitive to how \textit{Planck} and ACT data are combined, varying by $20\%$ across the likelihood combinations considered in Ref.~\cite{DESI:2025gwf}. In \cref{sec:additional_MCMC_results}, we present the \textit{Planck}+DESI+lens(PR3) analysis. This uses the \textit{Planck} PR3 CMB lensing reconstruction, denoted by \textbf{lens(PR3)}, to obtain constraints independent of ACT. The results are very similar to the \textit{Planck}+DESI analysis, indicating that the \textit{Planck} lensing likelihood adds little constraining power.

To account for the effects of $Y$ decay discussed in \cref{subsec:effects}, we introduce two additional likelihoods that are included in all MCMC analyses of the decay model. To constrain spectral distortions, we implement Gaussian likelihoods for $\mu$ and $y$, denoted collectively by \textbf{SD}. Each distribution has zero mean and $2 \sigma$ limits corresponding to the 95\% confidence intervals from COBE/FIRAS~\cite{Fixsen:1996nj},
\begin{subequations}
\label{eq:sd_bounds}
\begin{align}
    |\mu| &< \num{9.0e-5},
    \label{eq:mu_sd_bound}\\
    |y| &< \num{1.5e-5}.
    \label{eq:y_sd_bound}
\end{align}
\end{subequations}

We also use a Gaussian likelihood for the primordial deuterium abundance, denoted by \textbf{D/H}. Ref.~\cite{Cooke:2017cwo} gives $\num{e5} \mathrm{D/H} = \num{2.527 (0.030)}$~(68\%~CI), which only includes the observational D/H uncertainty. As in Refs.~\cite{Sobotka:2022vrr, Sobotka:2023bzr}, we broaden this constraint to account for the uncertainty in the BBN nuclear reaction rates. For the $d(p, \gamma)^3$He cross section in Ref.~\cite{Adelberger:2010qa}, the analysis of Ref.~\cite{Cooke:2017cwo} yields $100 \Omega_{\mathrm{b}} h^2 (\mathrm{BBN}) = \num{2.235(0.016)(0.033)}$~(68\%~CI), where the second error term represents the uncertainty in the BBN nuclear reaction rates. Combining both uncertainties in quadrature and converting to a baryon-to-photon ratio with $\num{e10} \eta = (\num{273.78(0.18)}) \Omega_{\mathrm{b}} h^2$~\cite{Steigman:2006nf, Cooke:2016rky} gives $\num{e10} \eta_{\mathrm{BBN}} = \num{6.119(0.100)}$~(68\%~CI). With our modified version of \texttt{PArthENoPE}, we find $\mathrm{D/H} \propto 1 / (\eta_{\mathrm{BBN}})^{1.65}$ for the $\eta_{\mathrm{BBN}}$ relevant to this work, $5.80 < \num{e10} \eta_{\mathrm{BBN}} < 6.88$. We propagate the $\eta_{\mathrm{BBN}}$ uncertainty into an effective D/H uncertainty using $\sigma_{\mathrm{D/H}} = 1.65 (\sigma_{\eta_{\mathrm{BBN}}} / \eta_{\mathrm{BBN}}) \mathrm{D/H}$, which sets the $1 \sigma$ width of our D/H likelihood,
\begin{equation}
    \num{e5} \mathrm{D/H} = \num{2.527(0.068)}.
    \label{eq:dh_bounds}
\end{equation}

\subsection{Numerical implementation}
\label{subsec:numerical_implementation}

Cosmological calculations are performed with a modified version of the Boltzmann solver \texttt{CLASS v3.1.0}\footnote{\url{https://github.com/SebastianJFigueroa/CLASS_YPlus}}~\cite{Blas:2011rf} and \texttt{HMcode} for nonlinear corrections~\cite{Mead:2016zqy}. For MCMC sampling, we employ a modified version of \texttt{MontePython v3.6.1}\footnote{\url{https://github.com/SebastianJFigueroa/MontePython_NuTwo}}~\cite{Audren:2012wb, Brinckmann:2018cvx} with a Metropolis--Hastings algorithm, and we use \texttt{GetDist} for MCMC post-processing and plotting~\cite{Lewis:2019xzd}. We assess convergence with the Gelman--Rubin diagnostic~\cite{Gelman:1992zz}, requiring $|R - 1| < 0.01$ for $\Lambda$CDM runs and $|R - 1| < 0.04$ for $Y$-decay runs, where the relaxed threshold reflects the non-Gaussian posteriors of the decay parameters. See \cref{tab:mcmc_priors} for a list of relevant priors.

We analyze several combinations of datasets under the DO, in which the sum of the neutrino masses is sampled with a flat prior of $\sum m_{\nu} \, [\unit{\electronvolt}] \in [0, \infty)$. The DO yields tighter bounds than the NO and IO unless the prior on $\sum m_{\nu}$ is consistent with the measured neutrino mass splittings~\cite{Archidiacono:2020dvx, Herold:2024nvk}. Some studies interpret recent cosmological results as favoring the NO~\cite{Long:2017dru, Jimenez:2022dkn}, while others trace this preference to parametrization and prior selection~\cite{RoyChoudhury:2019hls, Gariazzo:2022ahe}. Bayesian inferences of the neutrino masses are nonetheless sensitive to these sampling choices, particularly when the physical orderings are imposed~\cite{Gerbino:2016ehw, Heavens:2018adv, Gariazzo:2018meg, Mahony:2019fyb}. For the \textit{Planck}+DESI+lens(PR4+ACT) dataset, we therefore also consider the NO and IO. When using the NuFIT 6.0 results~\cite{Esteban:2024eli}, the solar splitting is given by $\Delta m_{\mathrm{sol}}^2 \equiv m_2^2 - m_1^2 = \qty{7.49(0.19:0.19)e-5}{\electronvolt\squared}$~(68\%~CI) in both orderings. The atmospheric splitting is defined so that its absolute value always parameterizes the mass-squared difference between the lightest and heaviest neutrino~\cite{Gonzalez-Garcia:2014bfa}. In the NO, where $m_1 < m_2 \ll m_3$, this yields $\Delta m_{\mathrm{atm}}^2 \equiv m_3^2 - m_1^2 = \qty{+2.513(0.021:0.019)e-3}{\electronvolt\squared}$~(68\%~CI). In the IO, where $m_3 \ll m_1 < m_2$, we instead have $\Delta m_{\mathrm{atm}}^2 \equiv m_3^2 - m_2^2 = \qty{-2.484(0.020:0.020)e-3}{\electronvolt\squared}$~(68\%~CI).

For our NO and IO analyses, we first employ the default \texttt{MontePython} approach, sampling $\sum m_{\nu}$ and using fixed mass splittings to derive $m_1$, $m_2$, and $m_3$, which are then passed as inputs to \texttt{CLASS}. We impose priors of $\sum m_{\nu} \, [\unit{\electronvolt}] \in [0.059, \infty)$ and $\sum m_{\nu} \, [\unit{\electronvolt}] \in [0.099, \infty)$ for the NO and IO, respectively. To incorporate the uncertainties in the mass splittings, we modify \texttt{MontePython} to sample $\Delta m_{\mathrm{sol}}^2$ and $\Delta m_{\mathrm{atm}}^2$ directly. We then conduct an MCMC analysis that varies the lightest neutrino mass with the prior $m_0 \, [\unit{\electronvolt}] \in [0, \infty)$, where $m_0 = m_1$ in the NO and $m_0 = m_3$ in the IO. With the NuFIT 6.0 constraints~\cite{Esteban:2024eli}, we employ Gaussian likelihoods for the mass splittings and enforce a prior of $\smash{\Delta m_{\mathrm{sol}}^2 \, [\unit{\electronvolt\squared}] \in [0, \infty)}$ in both orderings, $\Delta m_{\mathrm{atm}}^2 \, [\unit{\electronvolt\squared}] \in [0, \infty)$ in the NO, and $\Delta m_{\mathrm{atm}}^2 \, [\unit{\electronvolt\squared}] \in (-\infty, 0]$ in the IO.

\begin{table}
    \caption{MCMC priors for the neutrino mass orderings, $Y$-decay parameters, and $T_{\nu}$ in our model-agnostic analysis.}
    \label{tab:mcmc_priors}
    \begin{ruledtabular}
    \begin{tabular}{ll}
        \multicolumn{2}{l}{\textbf{DO}}\\
        $\sum m_{\nu} \, [\unit{\electronvolt}]$
        & $[0, \infty)$\\
        \addlinespace[1.5 pt]
        
        \multicolumn{2}{l}{\textbf{NO, IO (fixed)}}\\
        $\sum m_{\nu} \, [\unit{\electronvolt}]$
        & $[0.059, \infty)_{\mathrm{NO}}$, $[0.099, \infty)_{\mathrm{IO}}$\\
        \addlinespace[1.5 pt]
        
        \multicolumn{2}{l}{\textbf{NO, IO (marg.)}}\\
        $m_0 \, [\unit{\electronvolt}]$
        & $[0, \infty)$\\
        $\Delta m_{\mathrm{sol}}^2 \, [\unit{\electronvolt\squared}]$
        & $[0, \infty)$\\
        $\Delta m_{\mathrm{atm}}^2 \, [\unit{\electronvolt\squared}]$
        & $[0, \infty)_{\mathrm{NO}}$, $(-\infty, 0]_{\mathrm{IO}}$\\
        \midrule
        
        \multicolumn{2}{l}{$\boldsymbol{Y}$ \textbf{Decay}}\\
        $f_{\gamma}$
        & $[0.01, 1]$\\
        $\log_{10}(\Gamma_Y / \unit{\per\second})$
        & $[-6.46, -4.42]$\\
        $R_{\Gamma}$ & $[0, 0.07]$\\
        \addlinespace[1.5 pt]
    
        \multicolumn{2}{l}{\textbf{Varying} $\boldsymbol{T_{\nu}}$}\\
        $T_{\nu} \, [\unit{\electronvolt}]$
        & $[0, \infty)$
    \end{tabular}
    \end{ruledtabular}
\end{table}

We use flat priors on the six base cosmological parameters, $\{\omega_{\mathrm{b}}, \omega_{\mathrm{cdm}}, h, A_{\mathrm{s}}, n_{\mathrm{s}}, \tau_{\mathrm{reio}}\}$. Priors on the decay parameters are informed by Ref.~\cite{Sobotka:2022vrr}, which constrained the $Y$-decay scenario using the \textit{Planck}+SD+D/H dataset and one massive neutrino species with $m_{\nu} = \qty{0.06}{\electronvolt}$. That analysis yielded $R_{\Gamma} < 0.0235$~(95\%~CrI), so we adopt $R_{\Gamma} \in [0, 0.07]$ as a conservative prior. We aim to probe the neutrino-mass bound under changes to the neutrino abundance, which in the $Y$-decay model occurs when $f_{\gamma} > 0$. Allowing arbitrarily small values of $f_{\gamma}$ leads to volume effects, since $\Gamma_Y$ is no longer constrained by spectral-distortion bounds as $f_{\gamma} \rightarrow 0$. We therefore enforce a lower bound on $f_{\gamma}$ by using a prior of $f_{\gamma} \in [0.01, 1]$. With this prior, the COBE/FIRAS limits on spectral distortions exclude decay scenarios with $T_{\mathrm{RH}} \lesssim \qty{9.5e-4}{\mega\electronvolt}$. We also enforce that $Y$ decay begins after BBN has completed at $T \approx \qty{0.01}{\mega\electronvolt}$. These two constraints are implemented as a logarithmic prior on the decay rate, $\log_{10}(\Gamma_Y / \unit{\per\second}) \in [-6.46, -4.42]$. Finally, in our model-agnostic analysis where we vary $T_{\nu}$ directly, we adopt a flat prior of $T_{\nu} \, [\unit{\electronvolt}] \in [0, \infty)$.

\section{Results}
\label{sec:results}

\begin{figure}
    \includegraphics[width = \columnwidth]{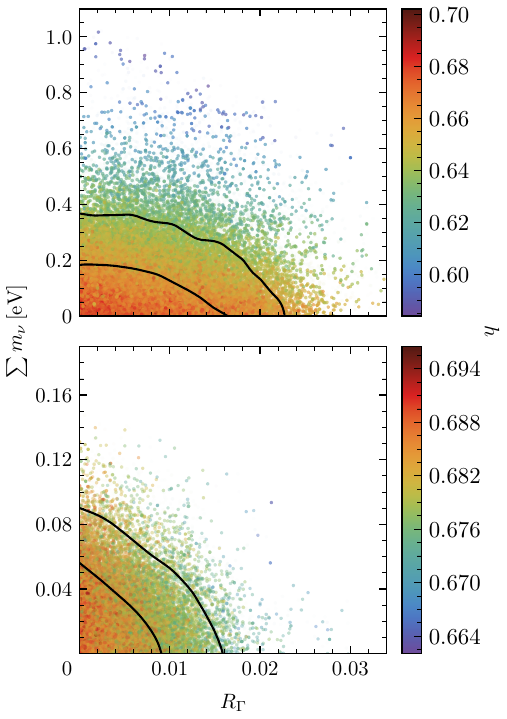}
    \caption{\textbf{Upper:} Accepted MCMC samples in the $R_{\Gamma}$--$\sum m_{\nu}$ plane using the DO approximation and the \textit{Planck}+SD+D/H dataset. We color each point by its value of $h$ and set $f_{\gamma} = 1$ to maximize the reduction in the neutrino abundance. Increasing either $R_{\Gamma}$ or $\sum m_{\nu}$ drives $h$ into a region incompatible with primary CMB data from \textit{Planck}. Reducing the neutrino abundance through photon injection therefore does not relax the bound on $\sum m_{\nu}$. \textbf{Lower:} As in the upper panel, but with the \textit{Planck}+DESI+lens(PR4+ACT)+SD+D/H dataset. Increasing either $\sum m_{\nu}$ or $R_{\Gamma}$ continues to push the posterior to lower $h$ and correspondingly larger $\Omega_{\mathrm{m}}$, opposite to the DESI preference. As a result, the inclusion of DESI data further restricts photon injection and the bound on $\sum m_{\nu}$.}
    \label{fig:m_tot_versus_max_ratio}
\end{figure}

\begin{table*}
    \caption{Marginalized constraints on the decay and cosmological parameters under the DO, quoted at 95\% credibility. For quantities common to both models, the upper entry in regular text shows the $\Lambda$CDM result, while the lower entry in boldface gives the $Y$-decay constraint. The $\Lambda$CDM fits use the same cosmological datasets as the $Y$-decay analyses, but omit the SD and D/H likelihoods. We use an ellipsis to denote cases in which the marginalized posterior does not yield a 95\% credible interval.}
    \label{tab:all_datasets_do}
    \begin{ruledtabular}
    \begin{tabular}{lccc}
        {}
        &
        \begin{tabular}{c}
            \makebox[\widthof{\textit{Planck}+DESI+lens(PR4+ACT)}][c]{\textit{Planck}}\\
            \textbf{+SD+D/H}
        \end{tabular}
        &
        \begin{tabular}{c}
            \makebox[\widthof{\textit{Planck}+DESI+lens(PR4+ACT)}][c]{\textit{Planck}+DESI}\\
            \textbf{+SD+D/H}
        \end{tabular}
        &
        \begin{tabular}{c}
            \textit{Planck}+DESI+lens(PR4+ACT)\\
            \textbf{+SD+D/H}
        \end{tabular}\\
        \midrule
        
        $f_{\gamma}$
        &
        $\cdots$
        &
        $\cdots$
        &
        $\cdots$\\
        \addlinespace[1.5 pt]
        
        $\log_{10}(\Gamma_Y / \unit{\per\second})$
        &
        $\mathbf{> -6.13}$
        &
        $\mathbf{> -6.15}$
        &
        $\mathbf{> -6.17}$\\
        \addlinespace[1.5 pt]
        
        $R_{\Gamma}$
        &
        $\mathbf{< 0.0230}$
        &
        $\mathbf{< 0.0249}$
        &
        $\mathbf{< 0.0232}$\\
        \midrule
        
        $\sum m_{\nu} \, [\unit{\electronvolt}]$
        &
        \begin{tabular}{c}
            $< 0.250$\\
            $\mathbf{< 0.272}$
        \end{tabular}
        &
        \begin{tabular}{c}
            $< 0.0792$\\
            $\mathbf{< 0.0819}$
        \end{tabular}
        &
        \begin{tabular}{c}
            $< 0.0691$\\
            $\mathbf{< 0.0710}$
        \end{tabular}\\
        \addlinespace[1.5 pt]
        
        $100 \omega_{\mathrm{b}}$
        &
        \begin{tabular}{c}
            $\num{2.234(0.030:0.030)}$\\
            $\mathbf{\num{2.232(0.032:0.031)}}$
        \end{tabular}
        &
        \begin{tabular}{c}
            $\num{2.251(0.025:0.025)}$\\
            $\mathbf{\num{2.249(0.026:0.025)}}$
        \end{tabular}
        &
        \begin{tabular}{c}
            $\num{2.253(0.025:0.025)}$\\
            $\mathbf{\num{2.252(0.025:0.025)}}$
        \end{tabular}\\
        \addlinespace[1.5 pt]
        
        $\omega_{\mathrm{cdm}}$
        &
        \begin{tabular}{c}
            $\num{0.1203(0.0027:0.0027)}$\\
            $\mathbf{\num{0.1209(0.0035:0.0035)}}$
        \end{tabular}
        &
        \begin{tabular}{c}
            $\num{0.1180(0.0014:0.0015)}$\\
            $\mathbf{\num{0.1190(0.0038:0.0028)}}$
        \end{tabular}
        &
        \begin{tabular}{c}
            $\num{0.1181(0.0013:0.0014)}$\\
            $\mathbf{\num{0.1191(0.0034:0.0026)}}$
        \end{tabular}\\
        \addlinespace[1.5 pt]
        
        $h$
        &
        \begin{tabular}{c}
            $\num{0.671(0.018:0.022)}$\\
            $\mathbf{\num{0.671(0.023:0.026)}}$
        \end{tabular}
        &
        \begin{tabular}{c}
            $\num{0.6858(0.0062:0.0063)}$\\
            $\mathbf{\num{0.688(0.012:0.010)}}$
        \end{tabular}
        &
        \begin{tabular}{c}
            $\num{0.6859(0.0060:0.0060)}$\\
            $\mathbf{\num{0.688(0.012:0.010)}}$
        \end{tabular}\\
        \addlinespace[1.5 pt]
        
        $\ln(10^{10} A_{\mathrm{s}})$
        &
        \begin{tabular}{c}
            $\num{3.045(0.032:0.031)}$\\
            $\mathbf{\num{3.046(0.033:0.031)}}$
        \end{tabular}
        &
        \begin{tabular}{c}
            $\num{3.046(0.033:0.031)}$\\
            $\mathbf{\num{3.048(0.035:0.032)}}$
        \end{tabular}
        &
        \begin{tabular}{c}
            $\num{3.053(0.027:0.025)}$\\
            $\mathbf{\num{3.054(0.027:0.026)}}$
        \end{tabular}\\
        \addlinespace[1.5 pt]
        
        $n_{\mathrm{s}}$
        &
        \begin{tabular}{c}
            $\num{0.9643(0.0088:0.0087)}$\\
            $\mathbf{\num{0.964(0.0110:0.0099)}}$
        \end{tabular}
        &
        \begin{tabular}{c}
            $\num{0.9699(0.0068:0.0068)}$\\
            $\mathbf{\num{0.9710(0.0084:0.0079)}}$
        \end{tabular}
        &
        \begin{tabular}{c}
            $\num{0.9705(0.0066:0.0066)}$\\
            $\mathbf{\num{0.9715(0.0084:0.0076)}}$
        \end{tabular}\\
        \addlinespace[1.5 pt]
        
        $\tau_{\mathrm{reio}}$
        &
        \begin{tabular}{c}
            $\num{0.054(0.016:0.015)}$\\
            $\mathbf{\num{0.054(0.016:0.015)}}$
        \end{tabular}
        &
        \begin{tabular}{c}
            $\num{0.057(0.016:0.015)}$\\
            $\mathbf{\num{0.057(0.016:0.015)}}$
        \end{tabular}
        &
        \begin{tabular}{c}
            $\num{0.060(0.015:0.014)}$\\
            $\mathbf{\num{0.059(0.015:0.014)}}$
        \end{tabular}\\
    \end{tabular}
    \end{ruledtabular}
\end{table*}

As discussed in \cref{sec:neutrino_abundance}, decreasing the neutrino abundance does not automatically relax the cosmological bound on $\sum m_{\nu}$, as evident in the upper panel of \cref{fig:m_tot_versus_max_ratio}. This plot shows results from an MCMC analysis of the $Y$-decay model using the \textit{Planck}+SD+D/H dataset, with $\sum m_{\nu}$ varied under the DO. We enforce decay exclusively into photons by setting $f_{\gamma} = 1$, which yields the maximum reduction in the neutrino abundance for a given value of $R_{\Gamma}$. Each point represents an accepted MCMC sample colored by its value of $h$, which we plot in the $R_{\Gamma}$--$\sum m_{\nu}$ plane. The largest accepted values of $\sum m_{\nu}$ occur at the smallest values of $R_{\Gamma}$, demonstrating that photon-injection scenarios do not make heavier neutrinos more consistent with CMB observations. In \cref{sec:neutrino_abundance}, we showed that reducing $N_{\mathrm{eff}}$ and increasing $\sum m_{\nu}$ both decrease the value of $h$ needed to keep $\theta_{\mathrm{s}}$ fixed. When both changes occur simultaneously, the required value of $h$ rapidly becomes too small to be accommodated by \textit{Planck} constraints. The reduction in $h$ is even more disfavored by DESI data, as seen in the lower panel of \cref{fig:m_tot_versus_max_ratio}, which shows analogous MCMC results for the $f_{\gamma} = 1$ decay scenario with the \textit{Planck}+DESI+lens(PR4+ACT)+SD+D/H dataset. For fixed physical matter density, decreasing $h$ increases $\Omega_{\mathrm{m}}$, in strong tension with the DESI preference for smaller values of $\Omega_{\mathrm{m}}$ than those favored by the CMB.

\begin{figure}[!t]
    \includegraphics[width = \columnwidth]{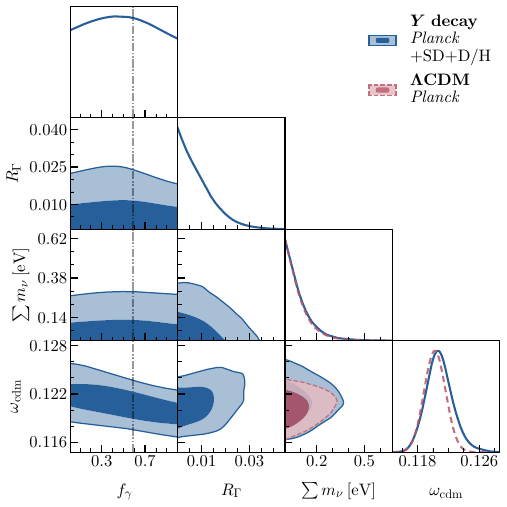}
    \caption{Marginalized posteriors for the $Y$-decay model and $\Lambda$CDM using the DO approximation and the \textit{Planck} dataset. The vertical line in the first column denotes $f_{\gamma} = 0.5913$, where the shallow peak in the $f_{\gamma}$ posterior reflects decays that maintain  $N_{\mathrm{eff}} = 3.044$. The slight preference for $f_{\gamma} \lesssim 0.59$ corresponds to decays dominated by DR injection, yielding $N_{\mathrm{eff}} > 3.044$. Increasing $N_{\mathrm{eff}}$ relative to $\Lambda$CDM reduces the effect of massive neutrinos on $\theta_{\mathrm{s}}$ and permits a larger $\omega_{\mathrm{cdm}}$ while preserving the observed MR equality scale.}
    \label{fig:triangle_planck}
    \vskip-\textfloatsep
\end{figure}

\begin{figure*}[t]
    \includegraphics[width = \textwidth]{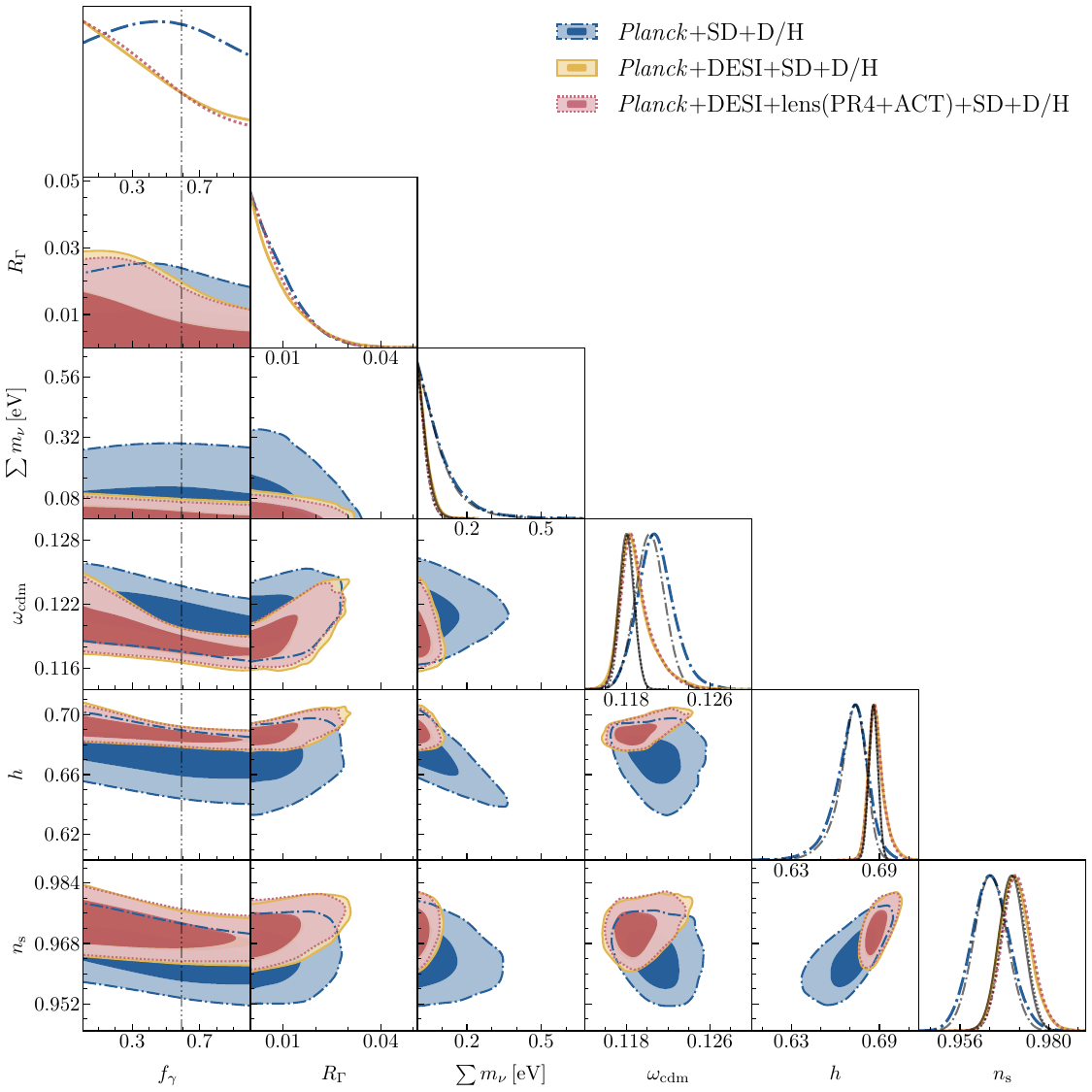}
    \caption{Summary of $Y$-decay DO posteriors for the three principal datasets, with $\Lambda$CDM posteriors shown on the diagonal in gray with reduced line width. The vertical line in the first column denotes $f_{\gamma} = 0.5913$. In the \textit{Planck} analysis, the marginal preference for $f_{\gamma} \lesssim 0.59$ reflects decays that yield $N_{\mathrm{eff}} > 3.044$, which permit higher $\omega_{\mathrm{cdm}}$ while preserving the MR equality scale. Adding DESI restricts the increase in the late-time matter density but strengthens the preference for $f_{\gamma} \lesssim 0.59$, as the increase in $N_{\mathrm{eff}}$ favors larger $h$ and smaller $\Omega_{\mathrm{m}}$. The larger $N_{\mathrm{eff}}$ and $Y_{\mathrm{He}}$ in these decay scenarios also enhance Silk damping and favor larger $n_{\mathrm{s}}$. Including ACT lensing maintains the strong preference for $f_{\gamma} \lesssim 0.59$ while slightly tightening all constraints.}
    \label{fig:triangle_principal_data}
\end{figure*}

All DO parameter constraints are collected in \cref{tab:all_datasets_do}, and the complete posterior distributions are shown in \cref{fig:triangle_all_data} in \cref{sec:additional_MCMC_results}. We first consider the \textit{Planck} results in \cref{fig:triangle_planck}, where the neutrino-mass bound weakens from $\sum m_{\nu} < \qty{0.250}{\electronvolt}$~(95\%~CrI;~DO) in $\Lambda$CDM to $\sum m_{\nu} < \qty{0.272}{\electronvolt}$~(95\%~CrI;~DO) with $Y$ decay. In Ref.~\cite{Sobotka:2022vrr}, an analysis of $Y$ decay using the same dataset but with fixed $\sum m_{\nu}$ yielded a nearly flat posterior for $f_{\gamma}$, with a shallow peak at $f_{\gamma} \approx 0.59$ reflecting decays that preserve $N_{\mathrm{eff}} = 3.044$ through recombination. While we recover this feature, we also obtain a slight preference for $f_{\gamma} \lesssim 0.59$ by marginalizing over $\sum m_{\nu}$. For decay scenarios with $f_{\gamma} < 0.5913$, the increase in $N_{\mathrm{eff}}$ reduces the change in $\theta_{\mathrm{s}}$ from massive neutrinos, thereby opening a larger volume of acceptable parameter space. The increased radiation density also permits a larger $\omega_{\mathrm{cdm}}$ while preserving the observed MR equality scale, reflected in the broadening of the upper tail of the $\omega_{\mathrm{cdm}}$ posterior. This increase in late-time matter density enhances CMB lensing, partially offsetting the suppression from neutrino free streaming and allowing larger neutrino masses.

The posteriors in \cref{fig:triangle_principal_data} show how parameter degeneracies and bounds change with the addition of DESI BAO and ACT CMB lensing data to the primary \textit{Planck} dataset. All two-dimensional contours correspond to $Y$-decay chains, and one-dimensional posteriors for $\Lambda$CDM are shown along the diagonal for comparison. For the \textit{Planck}+DESI dataset, the upper bound on the neutrino masses is relaxed from $\sum m_{\nu} < \qty{0.0792}{\electronvolt}$~(95\%~CrI;~DO) in $\Lambda$CDM to $\sum m_{\nu} < \qty{0.0819}{\electronvolt}$~(95\%~CrI;~DO) in the $Y$-decay scenario. Adding DESI to the \textit{Planck} CMB data substantially tightens the neutrino-mass bounds in both $\Lambda$CDM and the $Y$-decay model, while yielding a stronger preference for the $f_{\gamma} \lesssim 0.59$ branch. The increased radiation density enhances the expansion rate before recombination and decreases $r_{\mathrm{s}}$. Preserving $\theta_{\mathrm{s}}$ then requires $r_{\ast}$ to decrease, which is achieved by increasing $h$. The photon diffusion length instead scales as $r_{\mathrm{D}} \propto H^{-1 / 2}$, so $r_{\mathrm{D}}$ shrinks less than $r_{\mathrm{s}}$ for an equivalent increase in the pre-recombination expansion rate. Decays that raise $N_{\mathrm{eff}}$ therefore yield a larger angular damping scale $\theta_{\mathrm{D}} \equiv r_{\mathrm{D}} / r_{\ast}$ for the same $\theta_{\mathrm{s}}$~\cite{Hou:2011ec}. All decay scenarios increase the expansion rate and $\eta_{\mathrm{BBN}}$ relative to $\Lambda$CDM, which both produce a larger value of $Y_{\mathrm{He}}$. This reduces the number of free electrons at recombination, increasing the photon mean free path and further enhancing Silk damping. The resulting suppression of small-scale CMB power favors an increase in $n_{\mathrm{s}}$, which can help to preserve the CMB lensing potential as larger neutrino masses produce stronger small-scale suppression in $P(k)$.

\begin{figure}
    \includegraphics[width = \columnwidth]{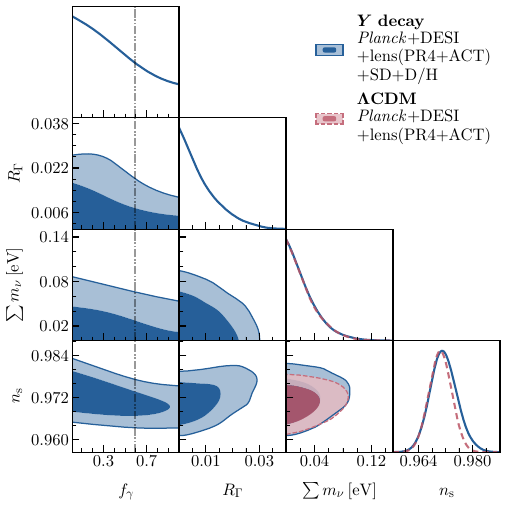}
    \caption{Marginalized posteriors for the $Y$-decay model and $\Lambda$CDM using the DO approximation and the \textit{Planck}+DESI+lens(PR4+ACT) dataset. The vertical line in the first column denotes $f_{\gamma} = 0.5913$. Compared to the \textit{Planck} analysis, the stronger preference for $f_{\gamma} \lesssim 0.59$ arises because decays that yield $N_{\mathrm{eff}} > 3.044$ favor higher $h$ and hence smaller $\Omega_{\mathrm{m}}$, as preferred by DESI. The increases in $N_{\mathrm{eff}}$ and $Y_{\mathrm{He}}$ also enhance Silk damping and favor larger $n_{\mathrm{s}}$. A bluer tilt partially compensates the small-scale suppression in $P(k)$ from neutrino free streaming, helping maintain the same lensing potential for slightly larger values of $\sum m_{\nu}$.}
    \label{fig:triangle_planck_desi_pr4_act}
\end{figure}

\begin{figure}
    \includegraphics[width = \columnwidth]{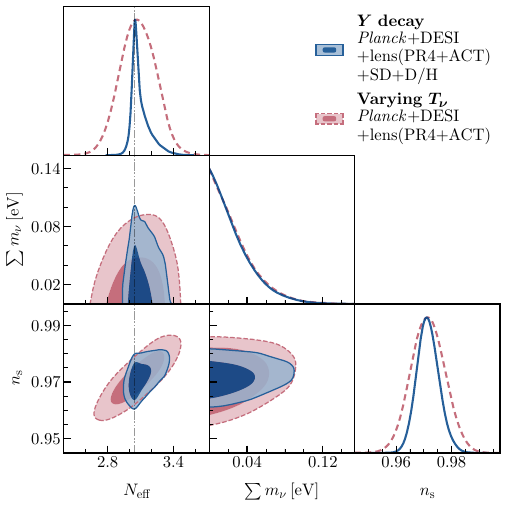}
    \caption{Marginalized posteriors for the $Y$-decay model and a model-agnostic analysis that varies $T_{\nu}$, using the DO approximation and the \textit{Planck}+DESI+lens(PR4+ACT) dataset. The vertical line in the first column denotes $N_{\mathrm{eff}} = 3.044$. Unlike in the generic $T_{\nu}$ analysis, $N_{\mathrm{eff}}>3.044$ in the $Y$-decay model does not require an increased neutrino abundance, because the additional radiation density can instead be supplied by DR. When varying $T_{\nu}$ directly, the largest neutrino masses still occur in the $N_{\mathrm{eff}} > 3.044$ region, despite the corresponding increase in the neutrino number density, as the larger $n_{\mathrm{s}}$ favored at higher $N_{\mathrm{eff}}$ partially offsets the increased small-scale suppression from a larger neutrino abundance.}
    \label{fig:triangle_model_agnostic_comparison}
\end{figure}

This slight $\sum m_{\nu}$--$n_{\mathrm{s}}$ correlation is visible in \cref{fig:triangle_planck_desi_pr4_act}, which highlights the MCMC results for the \textit{Planck}+DESI+lens(PR4+ACT) dataset under the DO approximation. The bound on the neutrino masses is relaxed from $\sum m_{\nu} < \qty{0.0691}{\electronvolt}$~(95\%~CrI;~DO) in $\Lambda$CDM to $\sum m_{\nu} < \qty{0.0710}{\electronvolt}$~(95\%~CrI;~DO) in the $Y$-decay model. We compute the Pearson and partial correlation coefficients for this dataset to identify the relevant degeneracies between the cosmological and decay parameters. \cref{sec:correlation_coefficients} contains the complete set of coefficients and a discussion of their implications. The familiar $\sum m_{\nu}$--$h$ anticorrelation gives the largest partial coefficient, $-0.349$. Increasing $\sum m_{\nu}$ lowers the $h$ required to preserve $\theta_{\mathrm{s}}$, so the higher $h$ favored by $f_{\gamma} < 0.5913$ decay scenarios provides additional room for larger neutrino masses. The positive $\sum m_{\nu}$--$n_{\mathrm{s}}$ correlation has the next-largest partial coefficient, 0.279. Decays with $f_{\gamma} < 0.5913$ favor a bluer tilt, and heavier neutrinos offset the resulting increase in small-scale power and help to preserve the observed CMB damping tail.

\begin{table*}
    \caption{Marginalized constraints on the decay and cosmological parameters under the NO and IO, quoted at 95\% credibility. DO results are included for comparison. All analyses use the \textit{Planck}+DESI+lens(PR4+ACT)+SD+D/H dataset, with the SD and D/H likelihoods omitted in the $\Lambda$CDM fits. Regular text shows the $\Lambda$CDM result, while entries in boldface give the general constraint for $Y$ decay. Quantities in italics are the $Y$-decay results with $f_{\gamma} = 1$, representing the tightest bounds on $\sum m_{\nu}$ in the decay scenario. We use a dash for entries that are not applicable and an overbar to denote a fixed parameter. In the fixed method, $m_0$ is derived from $\sum m_{\nu}$, while in the marginalized method, $\sum m_{\nu}$ is derived from $m_0$ and the mass splittings.}
    \label{tab:full_dataset_do_no_io}
    \begin{ruledtabular}
    \begin{tabular}{lccccc}
        {}
        &
        \begin{tabular}{c}
            \makebox[\widthof{$\mathbf{\num{-2.484(0.041:0.041)}}$}][c]{DO}
        \end{tabular}
        &
        \begin{tabular}{c}
            \makebox[\widthof{$\mathbf{\num{-2.484(0.041:0.041)}}$}][c]{NO (fixed)}
        \end{tabular}
        &
        \begin{tabular}{c}
            \makebox[\widthof{$\mathbf{\num{-2.484(0.041:0.041)}}$}][c]{IO (fixed)}
        \end{tabular}
        &
        \begin{tabular}{c}
            \makebox[\widthof{$\mathbf{\num{-2.484(0.041:0.041)}}$}][c]{NO (marg.)}
        \end{tabular}
        &
        \begin{tabular}{c}
            \makebox[\widthof{$\mathbf{\num{-2.484(0.041:0.041)}}$}][c]{IO (marg.)}
        \end{tabular}\\
        \midrule
        
        $f_{\gamma}$
        &
        \begin{tabular}{c}
            $\cdots$\\
            $\overline{\mathit{1.0}}$
        \end{tabular}
        &
        \begin{tabular}{c}
            $\mathbf{< 0.856}$\\
            $\overline{\mathit{1.0}}$
        \end{tabular}
        &
        \begin{tabular}{c}
            $\mathbf{< 0.865}$\\
            $\overline{\mathit{1.0}}$
        \end{tabular}
        &
        \begin{tabular}{c}
            $\mathbf{< 0.862}$\\
            $\overline{\mathit{1.0}}$
        \end{tabular}
        &
        \begin{tabular}{c}
            $\mathbf{< 0.832}$\\
            $\overline{\mathit{1.0}}$
        \end{tabular}\\
        \addlinespace[1.5 pt]
        
        $\log_{10}(\Gamma_Y / \unit{\per\second})$
        &
        \begin{tabular}{c}
            $\mathbf{> -6.17}$\\
            $\mathit{> -6.10}$
        \end{tabular}
        &
        \begin{tabular}{c}
            $\mathbf{> -6.15}$\\
            $\mathit{> -6.09}$
        \end{tabular}
        &
        \begin{tabular}{c}
            $\mathbf{> -6.13}$\\
            $\mathit{> -6.11}$
        \end{tabular}
        &
        \begin{tabular}{c}
            $\mathbf{> -6.14}$\\
            $\mathit{> -6.14}$
        \end{tabular}
        &
        \begin{tabular}{c}
            $\mathbf{> -6.16}$\\
            $\mathit{> -6.12}$
        \end{tabular}\\
        \addlinespace[1.5 pt]
        
        $R_{\Gamma}$
        &
        \begin{tabular}{c}
            $\mathbf{< 0.0232}$\\
            $\mathit{< 0.0116}$
        \end{tabular}
        &
        \begin{tabular}{c}
            $\mathbf{< 0.0271}$\\
            $\mathit{< 0.0119}$
        \end{tabular}
        &
        \begin{tabular}{c}
            $\mathbf{< 0.0297}$\\
            $\mathit{< 0.0103}$
        \end{tabular}
        &
        \begin{tabular}{c}
            $\mathbf{< 0.0269}$\\
            $\mathit{< 0.0107}$
        \end{tabular}
        &
        \begin{tabular}{c}
            $\mathbf{< 0.0315}$\\
            $\mathit{< 0.0107}$
        \end{tabular}\\
        \midrule
        
        $\sum m_{\nu} \, [\unit{\electronvolt}]$
        &
        \begin{tabular}{c}
        $\llap{$\scriptstyle{0 \leq} \left\{\vphantom{\begin{array}{c}
            < 0.0691\\
            \mathbf{< 0.0710}\\
            \mathit{< 0.0652}
        \end{array}}\right.$}
        \begin{array}{@{}c@{}}
            < 0.0691\\
            \mathbf{< 0.0710}\\
            \mathit{< 0.0652}
        \end{array}$
        \end{tabular}
        &
        \begin{tabular}{c}
        $\llap{$\scriptstyle{0.059 \leq} \left\{\vphantom{\begin{array}{c}
            < 0.109\\
            \mathbf{< 0.113}\\
            \mathit{< 0.106}
        \end{array}}\right.$}
        \begin{array}{@{}c@{}}
            < 0.109\\
            \mathbf{< 0.113}\\
            \mathit{< 0.106}
        \end{array}$
        \end{tabular}
        &
        \begin{tabular}{c}
        $\llap{$\scriptstyle{0.099 \leq} \left\{\vphantom{\begin{array}{c}
            < 0.141\\
            \mathbf{< 0.145}\\
            \mathit{< 0.141}
        \end{array}}\right.$}
        \begin{array}{@{}c@{}}
            < 0.141\\
            \mathbf{< 0.145}\\
            \mathit{< 0.141}
        \end{array}$
        \end{tabular}
        &
        \begin{tabular}{c}
            $\num{0.073(0.029:0.015)}$\\
            $\mathbf{\num{0.075(0.037:0.017)}}$\\
            $\mathit{\num{0.074(0.034:0.016)}}$
        \end{tabular}
        &
        \begin{tabular}{c}
            $\num{0.112(0.025:0.014)}$\\
            $\mathbf{\num{0.113(0.032:0.015)}}$\\
            $\mathit{\num{0.111(0.024:0.013)}}$
        \end{tabular}\\
        \addlinespace[1.5 pt]
        
        $m_0 \, [\unit{\electronvolt}]$
        &
        \begin{tabular}{c}
            $< 0.0230$\\
            $\mathbf{< 0.0237}$\\
            $\mathit{< 0.0217}$
        \end{tabular}
        &
        \begin{tabular}{c}
            $< 0.0257$\\
            $\mathbf{< 0.0272}$\\
            $\mathit{< 0.0245}$
        \end{tabular}
        &
        \begin{tabular}{c}
            $< 0.0278$\\
            $\mathbf{< 0.0295}$\\
            $\mathit{< 0.0274}$
        \end{tabular}
        &
        \begin{tabular}{c}
            $< 0.0225$\\
            $\mathbf{< 0.0256}$\\
            $\mathit{< 0.0240}$
        \end{tabular}
        &
        \begin{tabular}{c}
            $< 0.0255$\\
            $\mathbf{< 0.0287}$\\
            $\mathit{< 0.0248}$
        \end{tabular}\\
        \addlinespace[1.5 pt]
        
        $10^5 \Delta m_{\mathrm{sol}}^2 \, [\unit{\electronvolt\squared}]$
        &
        --
        &
        $\overline{\num{7.49}}$
        &
        $\overline{\num{7.49}}$
        &
        \begin{tabular}{c}
            $\num{7.49(0.38:0.37)}$\\
            $\mathbf{\num{7.49(0.38:0.38)}}$\\
            $\mathit{\num{7.49(0.37:0.38)}}$
        \end{tabular}
        &
        \begin{tabular}{c}
            $\num{7.48(0.38:0.36)}$\\
            $\mathbf{\num{7.48(0.37:0.38)}}$\\
            $\mathit{\num{7.50(0.40:0.39)}}$
        \end{tabular}\\
        \addlinespace[1.5 pt]
        
        $10^3 \Delta m_{\mathrm{atm}}^2 \, [\unit{\electronvolt\squared}]$
        &
        --
        &
        $\overline{\num{+2.513}}$
        &
        $\overline{\num{-2.484}}$
        &
        \begin{tabular}{c}
            $\num{2.514(0.041:0.040)}$\\
            $\mathbf{\num{2.513(0.039:0.038)}}$\\
            $\mathit{\num{2.514(0.040:0.038)}}$
        \end{tabular}
        &
        \begin{tabular}{c}
            $\num{-2.483(0.040:0.039)}$\\
            $\mathbf{\num{-2.484(0.041:0.041)}}$\\
            $\mathit{\num{-2.483(0.039:0.040)}}$
        \end{tabular}\\
    \end{tabular}
    \end{ruledtabular}
\end{table*}

\begin{figure}
    \includegraphics[width = \columnwidth]{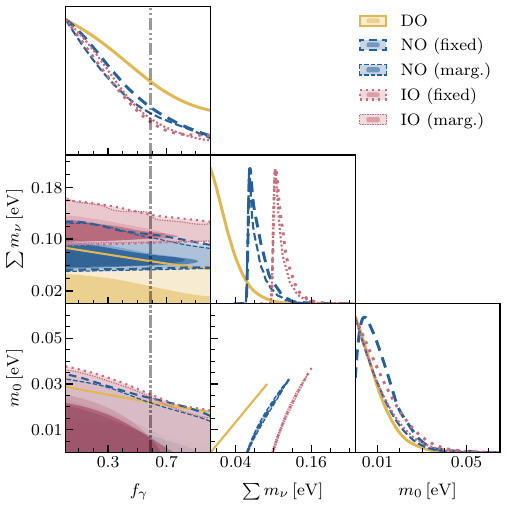}
    \caption{Marginalized posteriors for the $Y$-decay model using the \textit{Planck}+DESI+lens(PR4+ACT) dataset, comparing the DO with the NO and IO under two sampling prescriptions. The vertical line in the first column denotes $f_{\gamma} = 0.5913$. NO and IO analyses strengthen the preference for the $f_{\gamma} \lesssim 0.59$ branch, as their priors exclude the smaller values of $\sum m_{\nu}$ accessible in the DO. The increase in $N_{\mathrm{eff}}$ on this branch favors larger $n_{\mathrm{s}}$, helping accommodate the heavier neutrinos required by the physical orderings. The marginalized method yields tighter bounds than the fixed method due to different prior volumes from sampling $m_0$ rather than $\sum m_{\nu}$.}
    \label{fig:triangle_do_no_io_comparison}
\end{figure}

In \cref{fig:triangle_model_agnostic_comparison}, we present the results of a model-agnostic MCMC analysis in which we marginalize over $T_{\nu}$ directly, helping to assess whether the stringency of neutrino-mass bounds in the $Y$-decay scenario arises from constraints on the decay itself. We use the full \textit{Planck}+DESI+lens(PR4+ACT) dataset and plot the corresponding $Y$-decay posteriors for comparison. The Pearson and partial correlation coefficients for the generic $T_{\nu}$ analysis are given in \cref{sec:correlation_coefficients}. When directly varying $T_{\nu}$, $N_{\mathrm{eff}} > 3.044$ corresponds to a hotter neutrino population with a larger abundance compared to $\Lambda$CDM, while the same region in the $Y$-decay model can retain a reduced neutrino abundance, as the additional radiation density is supplied by DR. In both models, $N_{\mathrm{eff}} < 3.044$ corresponds to a reduced neutrino abundance relative to $\Lambda$CDM. By marginalizing over $T_{\nu}$, we find $\sum m_{\nu} < \qty{0.0724}{\electronvolt}$~(95\%~CrI;~DO) and $N_{\mathrm{eff}} = \num{3.07(0.31:0.32)}$~(95\%~CrI). For the direct-$T_{\nu}$ analysis, the high-$\sum m_{\nu}$ region in the two-dimensional $N_{\mathrm{eff}}$--$\sum m_{\nu}$ posterior extends furthest near $N_{\mathrm{eff}} \approx 3.2$, despite the accompanying increase in the relic-neutrino abundance relative to $\Lambda$CDM. As previously discussed, the increase in $N_{\mathrm{eff}}$ favors a larger $n_{\mathrm{s}}$, which can offset the stronger small-scale suppression of the lensing potential from the increased neutrino abundance. Furthermore, the strong partial correlation of 0.857 between $\sum m_{\nu}$ and $N_{\mathrm{eff}}$ reflects their opposing effects on $\theta_{\mathrm{s}}$, which allows both to increase while preserving the observed acoustic scale.

We obtain a similar but slightly tighter constraint of $N_{\mathrm{eff}} = \num{3.10(0.20:0.13)}$~(95\%~CrI) in the $Y$-decay model with photon and DR injection, where the neutrino-mass bound is $\sum m_{\nu} < \qty{0.0710}{\electronvolt}$~(95\%~CrI;~DO). Including the D/H likelihood severely constrains decay scenarios that yield $N_{\mathrm{eff}} < 3.044$ due to the change in $\eta_{\mathrm{BBN}}$ that arises from significant photon injection. The results in \cref{fig:triangle_model_agnostic_comparison} show how the high-$N_{\mathrm{eff}}$ tail opens parameter space beyond $\sum m_{\nu} \gtrsim \qty{0.06}{\electronvolt}$, while the $N_{\mathrm{eff}} < 3.044$ region does not. The accumulation at $N_{\mathrm{eff}} \approx 3.04$ in the two-dimensional $N_{\mathrm{eff}}$--$\sum m_{\nu}$ posterior is largely a volume effect from decays with small $R_{\Gamma}$, which give $N_{\mathrm{eff}} \simeq 3.04$ for all values of $f_{\gamma}$. Crucially, the $\sum m_{\nu}$ bound in the $Y$-decay scenario is only marginally tighter in the analysis where $T_{\nu}$ is varied directly, indicating that constraints on $Y$ decay do not drive the stringency of this limit. In both models, heavier neutrinos are accommodated by the $N_{\mathrm{eff}} > 3.044$ branch, which corresponds to a larger neutrino abundance in our model-agnostic analysis. Reducing $N_{\mathrm{eff}}$ therefore leads to tighter bounds on $\sum m_{\nu}$ in both the $Y$-decay scenario and when directly marginalizing over $T_{\nu}$.

To obtain neutrino-mass constraints that take the measured mass splittings into account, we perform both NO and IO analyses for the \textit{Planck}+DESI+lens(PR4+ACT) dataset combination using two sampling prescriptions. In the first, denoted as \textbf{fixed}, we sample the sum of the neutrino masses $\sum m_{\nu}$ and use fixed central values for the measured mass splittings to uniquely determine $m_1$, $m_2$, and $m_3$. In the second, denoted as \textbf{marginalized}, we sample the lightest neutrino mass $m_0$ and the two mass splittings, with the mass-splitting uncertainties incorporated through Gaussian likelihoods. Our complete results are presented in \cref{tab:full_dataset_do_no_io}, which shows that the physical orderings yield looser neutrino-mass bounds overall, primarily because the measured mass splittings impose a nonzero minimum value of $\sum m_{\nu}$. This difference is visible in \cref{fig:triangle_do_no_io_comparison}, which compares the $Y$-decay posteriors across the different mass orderings and sampling prescriptions. In both the fixed and marginalized methods, the physical orderings strengthen the preference for the $f_{\gamma} \lesssim 0.59$ branch relative to the DO, as their priors exclude the smaller values of $\sum m_{\nu}$ accessible in the DO. The larger neutrino masses sampled in the NO and IO analyses therefore further favor $Y$-decay scenarios with $f_{\gamma} \lesssim 0.59$, where the increase in $N_{\mathrm{eff}}$ supports larger $n_{\mathrm{s}}$ and partially offsets the stronger small-scale suppression of the lensing potential from heavier neutrinos.

Using the fixed sampling approach, we obtain constraints of $\sum m_{\nu} < \qty{0.113}{\electronvolt}$~(95\%~CrI;~NO$_{\mathrm{fixed}}$) and $\sum m_{\nu} < \qty{0.145}{\electronvolt}$~(95\%~CrI;~IO$_{\mathrm{fixed}}$) when allowing photon and DR injection. Both limits represent a relaxation of only $\qty{0.004}{\electronvolt}$ from the corresponding NO and IO bounds for a standard thermal history. Compared to the fixed method, the marginalized method yields slightly stricter neutrino-mass constraints in both the $Y$-decay model and $\Lambda$CDM. Allowing the $Y$ particles to decay into both photons and DR, the limit on the lightest neutrino mass tightens from $m_0 < \qty{0.0272}{\electronvolt}$~(95\%~CrI;~NO$_{\mathrm{fixed}}$) to $m_0 < \qty{0.0256}{\electronvolt}$~(95\%~CrI;~NO$_{\mathrm{marg.}}$) for the NO, and from a limit of $m_0 < \qty{0.0295}{\electronvolt}$~(95\%~CrI;~IO$_{\mathrm{fixed}}$) to $m_0 < \qty{0.0287}{\electronvolt}$~(95\%~CrI;~IO$_{\mathrm{marg.}}$) for the IO. As discussed in \cref{sec:prior_volume_effects}, the marginalized approach places relatively more prior weight at low values of $m_0$ and $\sum m_{\nu}$ than in the fixed approach, leading to stricter bounds on both parameters. For either sampling method, the NO and IO limits on the lightest neutrino mass relax by at most $\qty{0.0032}{\electronvolt}$ in the $Y$-decay analyses relative to $\Lambda$CDM. For all mass orderings, we also performed MCMC analyses of the $Y$-decay scenario with $f_{\gamma} = 1$, maximizing the reduction of the neutrino abundance for a given $R_{\Gamma}$. The resulting bounds on $\sum m_{\nu}$ and $m_0$ in \cref{tab:full_dataset_do_no_io} are consistently tighter than when DR is among the decay products, and in some cases even stricter than in $\Lambda$CDM, further underscoring the robustness of cosmological neutrino-mass bounds to photon injection.

\section{Summary and conclusions}
\label{sec:summary_and_conclusions}

In this work, we investigate how modifying the relic-neutrino abundance affects the cosmological limit on the sum of the neutrino masses. If cosmology only constrained the present-day energy density of neutrinos, $\rho_{\nu, 0}$, decreasing their number density relative to the standard $\Lambda$CDM prediction would relax the inferred neutrino-mass bound. We first consider a specific physical model that can reduce the neutrino abundance, consisting of massive $Y$ particles that decay into photons and DR between BBN and recombination, with a branching ratio that determines the fraction of the $Y$ particles that decay into photons. Photon injection lowers the neutrino abundance inferred from the observed temperature of the CMB and reduces $N_{\mathrm{eff}}$, while DR injection increases $N_{\mathrm{eff}}$ without changing the neutrino number density. For a given abundance of $Y$ particles, the neutrino density is maximally reduced if the $Y$ particles only decay into photons.

Our MCMC analyses employ primary CMB spectra from \textit{Planck}, CMB lensing reconstructions from \textit{Planck} and ACT, and BAO data from DESI. For the combined dataset, we obtain $\sum m_{\nu}<\qty{0.0652}{\electronvolt}$~(95\%~CrI;~DO) when the $Y$ particles decay exclusively into photons, which is even more restrictive than the constraint of $\sum m_{\nu} < \qty{0.0691}{\electronvolt}$~(95\%~CrI;~DO) for a standard thermal history. When DR injection is also allowed, this limit slightly relaxes to $\sum m_{\nu} < \qty{0.0710}{\electronvolt}$~(95\%~CrI;~DO). In decay scenarios with significant DR injection, the larger $N_{\mathrm{eff}}$ and $Y_{\mathrm{He}}$ relative to $\Lambda$CDM slightly increase the preferred value of $n_{\mathrm{s}}$, and a bluer tilt helps preserve the lensing potential as $\sum m_{\nu}$ increases. In contrast, reducing the neutrino abundance tightens the bound on $\rho_{\nu, 0}$ enough to yield a stronger limit on $\sum m_{\nu}$ because the accompanying decrease in $N_{\mathrm{eff}}$ increases $\theta_{\mathrm{s}}$. Since a larger value of $\rho_{\nu, 0}$ also increases $\theta_{\mathrm{s}}$, the CMB is even less tolerant of heavier neutrinos at smaller neutrino abundances.

We perform complementary MCMC analyses of the full dataset combination for the NO and IO, which yield slightly weaker constraints on the sum of the neutrino masses than in the DO due to the lower limits imposed by the measured mass splittings. The NO and IO analyses exhibit the same qualitative dependence on $N_{\mathrm{eff}}$; decay scenarios that are dominated by DR injection and yield $N_{\mathrm{eff}} > 3.044$ produce the most permissive neutrino-mass bounds. With fixed mass splittings, we obtain $\sum m_{\nu} < \qty{0.113}{\electronvolt}$~(95\%~CrI;~NO$_{\mathrm{fixed}}$) and $\sum m_{\nu} < \qty{0.145}{\electronvolt}$~(95\%~CrI;~IO$_{\mathrm{fixed}}$) for $Y$ particles that decay into both photons and DR. For both orderings, the upper limits on $\sum m_{\nu}$ are only $\qty{0.004}{\electronvolt}$ looser than the bounds obtained by assuming a standard thermal history.

In addition, we conduct NO and IO analyses that marginalize over the experimental uncertainties in the neutrino mass splittings. For these MCMC runs, we sample $m_0$ with a flat prior, which yields derived constraints of $\sum m_{\nu} = \qty{0.075(0.037:0.017)}{\electronvolt}$~(95\%~CrI;~NO$_{\mathrm{marg.}}$) and $\sum m_{\nu} = \qty{0.113(0.032:0.015)}{\electronvolt}$~(95\%~CrI;~IO$_{\mathrm{marg.}}$) when allowing both photon and DR injection from the decay of the $Y$ particles. The same posteriors give one-sided bounds of $\sum m_{\nu} < \qty{0.109}{\electronvolt}$~(95\%~CrI;~NO$_{\mathrm{marg.}}$) and $\sum m_{\nu} < \qty{0.143}{\electronvolt}$~(95\%~CrI;~IO$_{\mathrm{marg.}}$), which are slightly tighter than the corresponding limits in the fixed method that samples $\sum m_{\nu}$ with a flat prior. The Jacobian of the transformation between $m_0$ and $\sum m_{\nu}$ induces a non-flat prior on $\sum m_{\nu}$ in the marginalized approach, giving stronger neutrino-mass constraints. For the marginalized method, bounds on the lightest neutrino mass are $m_0 < \qty{0.0256}{\electronvolt}$~(95\%~CrI;~NO$_{\mathrm{marg.}}$) and $m_0 < \qty{0.0287}{\electronvolt}$~(95\%~CrI;~IO$_{\mathrm{marg.}}$) when the $Y$ particles decay into photons and DR. Relative to the $m_0$ bounds for a standard thermal history, these NO and IO constraints exhibit a maximum relaxation of only $\qty{0.0032}{\electronvolt}$.

Across all analyses of the $Y$-decay scenario, varying the neutrino abundance shifts the neutrino-mass constraints by at most $8.8\%$, with the largest shift occurring in the \textit{Planck}-only DO analysis of the $Y$-decay scenario. For the full dataset, the bound on $\sum m_{\nu}$ changes by a maximum of $5.6\%$. These shifts in the neutrino-mass limits due to $Y$ decay are notably smaller than those induced by changing the \textit{Planck} primary CMB likelihood used for parameter inference. When combined with the same DESI DR2 BAO data and lensing reconstruction from \textit{Planck} PR4 and ACT DR6 that we implement in this work, the use of the \texttt{LoLLiPoP-HiLLiPoP} likelihood rather than \texttt{CamSpec} increases the upper bound on $\sum m_{\nu}$ by $21\%$~\cite{Elbers:2025vlz}.

To test whether constraints on the decay scenario drive the robustness of our neutrino-mass bounds, we run an MCMC analysis that directly varies the C${\nu}$B temperature without assuming a specific physical model. Using all datasets, we obtain $\sum m_{\nu}<\qty{0.0724}{\electronvolt}$~(95\%~CrI;~DO) and $N_{\mathrm{eff}} = \num{3.07(0.31:0.32)}$~(95\%~CrI). This is in close agreement with the results of the $Y$-decay model with photon and DR injection, $\sum m_{\nu} < \qty{0.0710}{\electronvolt}$~(95\%~CrI;~DO) and $N_{\mathrm{eff}} = \num{3.10(0.20:0.13)}$~(95\%~CrI), showing that the stringency of our $\sum m_{\nu}$ bounds is not due to constraints on the decay model. As in the case of $Y$ decay, heavier neutrinos are permitted by increasing $N_{\mathrm{eff}}$, even as larger $N_{\mathrm{eff}}$ corresponds to a higher neutrino abundance in this analysis.

The primary CMB measurements from ACT DR6 favor a reduced neutrino abundance and yield a constraint of $N_{\mathrm{eff}} = \num{2.86(0.13)}$~(68\%~CrI) when combined with \textit{Planck} PR3 and \texttt{SRoll} primary CMB data, the \textit{Planck} PR4 and ACT DR6 lensing reconstruction, and DESI DR1 BAO \cite{AtacamaCosmologyTelescope:2025nti}. Entropy injection between neutrino decoupling and deuterium formation can produce such low values of $N_{\mathrm{eff}}$~\cite{Escudero:2026mgw}, but our results indicate that this would only tighten the cosmological bound on $\sum m_{\nu}$. Indeed, the ACT collaboration found that allowing $N_{\mathrm{eff}}$ to vary tightened its limit from $\sum m_{\nu} < \qty{0.089}{\electronvolt}$~(95\%~CrI;~DO) to $\sum m_{\nu} < \qty{0.073}{\electronvolt}$~(95\%~CrI;~DO)~\cite{AtacamaCosmologyTelescope:2025nti}. This preference for neutrino masses below the minimum value allowed in the NO could be alleviated by introducing dynamical dark energy. In the DESI DR2 analysis, the $\Lambda$CDM bound of $\sum m_{\nu} < \qty{0.0642}{\electronvolt}$~(95\%~CrI;~DO) is significantly loosened to $\sum m_{\nu} < \qty{0.163}{\electronvolt}$~(95\%~CrI;~DO) within the $w_0 w_a$CDM model, removing the tension with the lower limit from oscillation experiments~\cite{Elbers:2025vlz}.

By contributing to the radiation density at recombination and to the matter density today, the impact of massive neutrinos on the expansion history and structure formation is uniquely constrained by observations of the late-time universe and the CMB. Reducing the neutrino abundance tightens neutrino-mass limits, and only when additional DR compensates for the decrease in radiation density does the bound on the sum of the neutrino masses relax slightly. Changes to the neutrino abundance are therefore ineffective at resolving the emerging neutrino-mass tension, and any decrease in $N_{\mathrm{eff}}$ will only exacerbate the discrepancy between cosmological constraints on the neutrino masses and the measured mass splittings.

\begin{acknowledgments}

We thank Itamar Allali for introducing us to partial correlation coefficients. Our computational analyses employed the Longleaf cluster at the University of North Carolina~(UNC) at Chapel Hill. We acknowledge that UNC Chapel Hill is situated on the traditional homelands of the Occaneechi, Shakori, Eno, and Sissipahaw peoples. SJF, ACS, and ALE received support from the U.S. National Science Foundation~(NSF) under Award No. PHY-2310719. SJF was also supported by the Bahnson Fund at UNC Chapel Hill and the North Carolina Space Grant Graduate Research Fellowship under Cooperative Agreement No. 80NSSC25M7117 with the National Aeronautics and Space Administration. This work was performed in part at the Aspen Center for Physics, which is supported by NSF Award No. PHY-2210452.

\end{acknowledgments}

\appendix
\crefalias{section}{appendix}

\section{Correlation coefficients}
\label{sec:correlation_coefficients}

\begin{table*}
    \caption{Pearson correlation coefficients for the decay and cosmological parameters using the DO approximation and the \textit{Planck}+DESI+lens(PR4+ACT)+SD+D/H dataset. The magnitude of each coefficient indicates the strength of the overall linear relationship between the two parameters, while its sign indicates a positive or negative correlation.}
    \label{tab:pearson_correlations_y_decay}
    \begin{ruledtabular}
    \begin{tabular}{@{}l *{10}{S[table-format = 1.3]}@{}}
        {}
        &
        \multicolumn{1}{c}{\makebox[0pt][c]{$f_{\gamma}$}}
        &
        \multicolumn{1}{c}{\makebox[0pt][c]{$\log_{10}(\Gamma_Y / \unit{\per\second})$}}
        &
        \multicolumn{1}{c}{\makebox[0pt][c]{$R_{\Gamma}$}}
        &
        \multicolumn{1}{c}{\makebox[0pt][c]{$\sum m_{\nu} \, [\unit{\electronvolt}]$}}
        &
        \multicolumn{1}{c}{\makebox[0pt][c]{$\omega_{\mathrm{b}}$}}
        &
        \multicolumn{1}{c}{\makebox[0pt][c]{$\omega_{\mathrm{cdm}}$}}
        &
        \multicolumn{1}{c}{\makebox[0pt][c]{$h$}}
        &
        \multicolumn{1}{c}{\makebox[0pt][c]{$\ln(10^{10} A_{\mathrm{s}})$}}
        &
        \multicolumn{1}{c}{\makebox[0pt][c]{$n_{\mathrm{s}}$}}
        &
        \multicolumn{1}{c}{\makebox[0pt][c]{$\tau_{\mathrm{reio}}$}}\\
        \midrule
        
        $f_{\gamma}$
        & 1.000 & -0.097 & -0.330 & 0.121 & -0.236 & -0.676 & -0.659 & -0.182 & -0.352 & 0.002\\
        
        $\log_{10}(\Gamma_Y / \unit{\per\second})$
        & -0.097 & 1.000 & 0.214 & -0.067 & 0.052 & 0.174 & 0.161 & 0.109 & 0.019 & 0.075\\
        
        $R_{\Gamma}$
        & -0.330 & 0.214 & 1.000 & -0.094 & 0.148 & 0.685 & 0.579 & 0.213 & 0.284 & 0.034\\
        
        $\sum m_{\nu} \, [\unit{\electronvolt}]$
        & 0.121 & -0.067 & -0.094 & 1.000 & 0.137 & -0.225 & -0.148 & 0.166 & 0.135 & 0.187\\
        
        $\omega_{\mathrm{b}}$
        & -0.236 & 0.052 & 0.148 & 0.137 & 1.000 & 0.184 & 0.533 & 0.228 & 0.338 & 0.106\\
        
        $\omega_{\mathrm{cdm}}$
        & -0.676 & 0.174 & 0.685 & -0.225 & 0.184 & 1.000 & 0.607 & 0.135 & 0.221 & -0.097\\
        
        $h$
        & -0.659 & 0.161 & 0.579 & -0.148 & 0.533 & 0.607 & 1.000 & 0.314 & 0.601 & 0.127\\
        
        $\ln(10^{10} A_{\mathrm{s}})$
        & -0.182 & 0.109 & 0.213 & 0.166 & 0.228 & 0.135 & 0.314 & 1.000 & 0.206 & 0.924 \\
        
        $n_{\mathrm{s}}$
        & -0.352 & 0.019 & 0.284 & 0.135 & 0.338 & 0.221 & 0.601 & 0.206 & 1.000 & 0.177\\
        
        $\tau_{\mathrm{reio}}$
        & 0.002 & 0.075 & 0.034 & 0.187 & 0.106 & -0.097 & 0.127 & 0.924 & 0.177 & 1.000\\
    \end{tabular}
    \end{ruledtabular}
\end{table*}

\begin{table*}
    \caption{Partial correlation coefficients for the decay and cosmological parameters using the DO approximation and \textit{Planck}+DESI+lens(PR4+ACT)+SD+D/H dataset. The magnitude of each coefficient indicates the strength of the intrinsic linear relationship between the two parameters, while its sign denotes a positive or negative degeneracy. Unlike Pearson coefficients, partial coefficients remove the linear influence of the other parameters to isolate the relationship between each pair.}
    \label{tab:partial_correlations_y_decay}
    \begin{ruledtabular}
    \begin{tabular}{@{}l *{10}{S[table-format = 1.3]}@{}}
        {}
        &
        \multicolumn{1}{c}{\makebox[0pt][c]{$f_{\gamma}$}}
        &
        \multicolumn{1}{c}{\makebox[0pt][c]{$\log_{10}(\Gamma_Y / \unit{\per\second})$}}
        &
        \multicolumn{1}{c}{\makebox[0pt][c]{$R_{\Gamma}$}}
        &
        \multicolumn{1}{c}{\makebox[0pt][c]{$\sum m_{\nu} \, [\unit{\electronvolt}]$}}
        &
        \multicolumn{1}{c}{\makebox[0pt][c]{$\omega_{\mathrm{b}}$}}
        &
        \multicolumn{1}{c}{\makebox[0pt][c]{$\omega_{\mathrm{cdm}}$}}
        &
        \multicolumn{1}{c}{\makebox[0pt][c]{$h$}}
        &
        \multicolumn{1}{c}{\makebox[0pt][c]{$\ln(10^{10} A_{\mathrm{s}})$}}
        &
        \multicolumn{1}{c}{\makebox[0pt][c]{$n_{\mathrm{s}}$}}
        &
        \multicolumn{1}{c}{\makebox[0pt][c]{$\tau_{\mathrm{reio}}$}}\\
        \midrule
        
        $f_{\gamma}$
        & 1.000 & 0.006 & 0.464 & -0.131 & 0.228 & -0.546 & -0.452 & -0.101 & -0.030 & 0.098\\
        
        $\log_{10}(\Gamma_Y / \unit{\per\second})$
        & 0.006 & 1.000 & 0.101 & -0.024 & 0.007 & 0.028 & 0.052 & -0.037 & -0.084 & 0.060\\
        
        $R_{\Gamma}$
        & 0.464 & 0.101 & 1.000 & 0.157 & -0.265 & 0.575 & 0.401 & 0.113 & 0.022 & -0.095\\
        
        $\sum m_{\nu} \, [\unit{\electronvolt}]$
        & -0.131 & -0.024 & 0.157 & 1.000 & 0.252 & -0.184 & -0.349 & 0.151 & 0.279 & -0.095\\
        
        $\omega_{\mathrm{b}}$
        & 0.228 & 0.007 & -0.265 & 0.252 & 1.000 & 0.025 & 0.507 & 0.175 & -0.034 & -0.176\\
        
        $\omega_{\mathrm{cdm}}$
        & -0.546 & 0.028 & 0.575 & -0.184 & 0.025 & 1.000 & -0.052 & 0.220 & -0.062 & -0.255\\
        
        $h$
        & -0.452 & 0.052 & 0.401 & -0.349 & 0.507 & -0.052 & 1.000 & 0.151 & 0.497 & -0.106\\
        
        $\ln(10^{10} A_{\mathrm{s}})$
        & -0.101 & -0.037 & 0.113 & 0.151 & 0.175 & 0.220 & 0.151 & 1.000 & -0.261 & 0.952\\
        
        $n_{\mathrm{s}}$
        & -0.030 & -0.084 & 0.022 & 0.279 & -0.034 & -0.062 & 0.497 & -0.261 & 1.000 & 0.260 \\
        
        $\tau_{\mathrm{reio}}$
        & 0.098 & 0.060 & -0.095 & -0.095 & -0.176 & -0.255 & -0.106 & 0.952 & 0.260 & 1.000 \\
    \end{tabular}
    \end{ruledtabular}
\end{table*}

\begin{table*}
    \caption{Pearson correlation coefficients when varying the neutrino temperature generically, using the DO approximation and the \textit{Planck}+DESI+lens(PR4+ACT) dataset. Since $N_{\mathrm{eff}} \propto T_{\nu}^4$, the coefficients for $N_{\mathrm{eff}}$ closely trace the correlations of $T_{\nu}$.}
    \label{tab:pearson_correlations_model_agnostic}
    \begin{ruledtabular}
    \begin{tabular}{@{}l *{8}{S[table-format = 1.3]}@{}}
        {}
        &
        \multicolumn{1}{c}{\makebox[0pt][c]{$N_{\mathrm{eff}}$}}
        &
        \multicolumn{1}{c}{\makebox[0pt][c]{$\sum m_{\nu} \, [\unit{\electronvolt}]$}}
        &
        \multicolumn{1}{c}{\makebox[0pt][c]{$\omega_{\mathrm{b}}$}}
        &
        \multicolumn{1}{c}{\makebox[0pt][c]{$\omega_{\mathrm{cdm}}$}}
        &
        \multicolumn{1}{c}{\makebox[0pt][c]{$h$}}
        &
        \multicolumn{1}{c}{\makebox[0pt][c]{$\ln(10^{10} A_{\mathrm{s}})$}}
        &
        \multicolumn{1}{c}{\makebox[0pt][c]{$n_{\mathrm{s}}$}}
        &
        \multicolumn{1}{c}{\makebox[0pt][c]{$\tau_{\mathrm{reio}}$}}\\
        \midrule
        
        $N_{\mathrm{eff}}$
        & 1.000 & 0.071 & 0.670 & 0.961 & 0.937 & 0.352 & 0.787 & -0.140\\
        
        $\sum m_{\nu} \, [\unit{\electronvolt}]$
        & 0.071 & 1.000 & 0.175 & -0.005 & -0.042 & 0.227 & 0.186 & 0.200\\
        
        $\omega_{\mathrm{b}}$
        & 0.670 & 0.175 & 1.000 & 0.579 & 0.770 & 0.259 & 0.672 & -0.093\\
        
        $\omega_{\mathrm{cdm}}$
        & 0.961 & -0.005 & 0.579 & 1.000 & 0.834 & 0.296 & 0.669 & -0.190\\
        
        $h$
        & 0.937 & -0.042 & 0.770 & 0.834 & 1.000 & 0.338 & 0.828 & -0.121\\
        
        $\ln(10^{10} A_{\mathrm{s}})$
        & 0.352 & 0.227 & 0.259 & 0.296 & 0.338 & 1.000 & 0.320 & 0.812\\
        
        $n_{\mathrm{s}}$
        & 0.787 & 0.186 & 0.672 & 0.669 & 0.828 & 0.320 & 1.000 & -0.028\\
        
        $\tau_{\mathrm{reio}}$
        & -0.140 & 0.200 & -0.093 & -0.190 & -0.121 & 0.812 & -0.028 & 1.000\\
    \end{tabular}
    \end{ruledtabular}
\end{table*}

\begin{table*}
    \caption{Partial correlation coefficients when varying the neutrino temperature generically, using the DO approximation and the \textit{Planck}+DESI+lens(PR4+ACT) dataset. Since $N_{\mathrm{eff}} \propto T_{\nu}^4$, the coefficients for $N_{\mathrm{eff}}$ closely trace the correlations of $T_{\nu}$.}
    \label{tab:partial_correlations_model_agnostic}
    \begin{ruledtabular}
    \begin{tabular}{@{}l *{8}{S[table-format = 1.3]}@{}}
        {}
        &
        \multicolumn{1}{c}{\makebox[0pt][c]{$N_{\mathrm{eff}}$}}
        &
        \multicolumn{1}{c}{\makebox[0pt][c]{$\sum m_{\nu} \, [\unit{\electronvolt}]$}}
        &
        \multicolumn{1}{c}{\makebox[0pt][c]{$\omega_{\mathrm{b}}$}}
        &
        \multicolumn{1}{c}{\makebox[0pt][c]{$\omega_{\mathrm{cdm}}$}}
        &
        \multicolumn{1}{c}{\makebox[0pt][c]{$h$}}
        &
        \multicolumn{1}{c}{\makebox[0pt][c]{$\ln(10^{10} A_{\mathrm{s}})$}}
        &
        \multicolumn{1}{c}{\makebox[0pt][c]{$n_{\mathrm{s}}$}}
        &
        \multicolumn{1}{c}{\makebox[0pt][c]{$\tau_{\mathrm{reio}}$}}\\
        \midrule
        
        $N_{\mathrm{eff}}$
        & 1.000 & 0.857 & -0.715 & 0.979 & 0.950 & -0.009 & 0.041 & 0.010\\
        
        $\sum m_{\nu} \, [\unit{\electronvolt}]$
        & 0.857 & 1.000 & 0.735 & -0.830 & -0.891 & 0.113 & 0.171 & -0.069\\
        
        $\omega_{\mathrm{b}}$
        & -0.715 & 0.735 & 1.000 & 0.670 & 0.798 & 0.005 & -0.004 & -0.031\\
        
        $\omega_{\mathrm{cdm}}$
        & 0.979 & -0.830 & 0.670 & 1.000 & -0.900 & 0.093 & -0.044 & -0.102\\
        
        $h$
        & 0.950 & -0.891 & 0.798 & -0.900 & 1.000 & 0.093 & 0.154 & -0.079\\
        
        $\ln(10^{10} A_{\mathrm{s}})$
        & -0.009 & 0.113 & 0.005 & 0.093 & 0.093 & 1.000 & -0.170 & 0.929\\
        
        $n_{\mathrm{s}}$
        & 0.041 & 0.171 & -0.004 & -0.044 & 0.154 & -0.170 & 1.000 & 0.170\\
        
        $\tau_{\mathrm{reio}}$
        & 0.010 & -0.069 & -0.031 & -0.102 & -0.079 & 0.929 & 0.170 & 1.000\\
    \end{tabular}
    \end{ruledtabular}
\end{table*}

The Pearson and partial correlation coefficients for the \textit{Planck}+DESI+lens(PR4+ACT)+SD+D/H analysis using the DO approximation are reported in \cref{tab:pearson_correlations_y_decay,tab:partial_correlations_y_decay}. Pearson coefficients describe the marginalized linear relationship between each parameter pair, whereas partial coefficients remove the dependence on the remaining sampled parameters to isolate the residual pairwise correlation. In each case, the magnitude of the coefficient quantifies the strength of the relationship, while its sign indicates the direction of the degeneracy. See Ref.~\cite{Allali:2025yvp} for details on how these quantities are computed from the covariance matrices generated by our MCMC analyses.

Increasing $\sum m_{\nu}$ raises the value of $\theta_{\mathrm{s}}$, so maintaining the observed acoustic scale shifts the fit toward smaller $\omega_{\mathrm{cdm}}$ or $h$, leading to negative Pearson correlations of $\sum m_{\nu}$ with both parameters. Raising $f_{\gamma}$ lowers $N_{\mathrm{eff}}$ and also increases $\theta_{\mathrm{s}}$, producing similar compensating shifts and negative Pearson correlations with both $\omega_{\mathrm{cdm}}$ and $h$. The resulting shared dependence yields the positive Pearson correlation of $0.121$ between $\sum m_{\nu}$ and $f_{\gamma}$, as they vary in the same direction as the values of $\omega_{\mathrm{cdm}}$ and $h$ change. On the preferred $f_{\gamma} \lesssim 0.59$ branch, increasing $R_{\Gamma}$ results in a larger value of $N_{\mathrm{eff}}$ and decreases $\theta_{\mathrm{s}}$, requiring shifts toward larger $\omega_{\mathrm{cdm}}$ or $h$. Since larger $\sum m_{\nu}$ favors compensating changes in the opposite direction, these shared dependencies contribute to the slight Pearson anticorrelation of $-0.094$ between $\sum m_{\nu}$ and $R_{\Gamma}$.

After these mutual dependencies with $\omega_{\mathrm{cdm}}$ and $h$ are removed, the $\sum m_{\nu}$--$f_{\gamma}$ and $\sum m_{\nu}$--$R_{\Gamma}$ correlations switch sign, consistent with the compensation required to preserve $\theta_{\mathrm{s}}$. The negative $\sum m_{\nu}$--$f_{\gamma}$ correlation of $-0.131$ reflects the fact that increasing either parameter shifts $\theta_{\mathrm{s}}$ to larger values, so increasing one limits the extent to which the other can increase. The positive partial correlation of 0.157 between $\sum m_{\nu}$ and $R_{\Gamma}$ is consistent with the behavior on the preferred $f_{\gamma} \lesssim 0.59$ branch, where increasing $R_{\Gamma}$ raises $N_{\mathrm{eff}}$. This decreases $\theta_{\mathrm{s}}$, partially offsetting the increase in $\theta_{\mathrm{s}}$ induced by heavier neutrinos.

The strongest partial correlations involving $\sum m_{\nu}$ are $-0.349$ with $h$, $0.279$ with $n_{\mathrm{s}}$, and $0.252$ with $\omega_{\mathrm{b}}$. As discussed in \cref{sec:neutrino_abundance}, decreasing $h$ mitigates the impact that heavier neutrinos have on the distance to the CMB, but reducing $N_{\mathrm{eff}}$ only makes it more difficult to keep $\theta_{\mathrm{s}}$ fixed while increasing $\sum m_{\nu}$. Furthermore, DESI constraints on $\Omega_{\mathrm{m}}$ disfavor smaller values of $h$. A larger value of $\omega_{\mathrm{b}}$ can compensate for the effect of massive neutrinos on $\theta_{\mathrm{s}}$ by reducing the sound speed of the baryon-photon fluid, but $\omega_{\mathrm{b}}$ is tightly constrained by the relative heights of the even and odd acoustic peaks. The positive partial correlation between $\sum m_{\nu}$ and $n_{\mathrm{s}}$ provides the most effective mechanism for weakening the neutrino-mass bound. A bluer tilt can maintain the same lensing potential with heavier neutrinos, and the enhanced Silk damping in $Y$-decay scenarios with $N_{\mathrm{eff}} > 3.044$ makes larger $n_{\mathrm{s}}$ compatible with the observed damping tail. 

In \cref{tab:pearson_correlations_model_agnostic,tab:partial_correlations_model_agnostic}, we also provide the Pearson and partial correlation coefficients for the model-agnostic analysis in which we vary $T_{\nu}$ directly, using the DO approximation and the \textit{Planck}+DESI+lens(PR4+ACT) dataset. We include the coefficients for $N_{\mathrm{eff}}$, which closely trace the corresponding correlations of $T_{\nu}$. Since $N_{\mathrm{eff}} \propto T_{\nu}^4$, the signs of the correlation coefficients are preserved, while the nonlinear mapping produces small differences in magnitude. Unlike in the $Y$-decay analysis, where $N_{\mathrm{eff}}$ mediates the correlations between the decay parameters and $\sum m_{\nu}$, varying $T_{\nu}$ directly allows us to examine the relation between $\sum m_{\nu}$ and $N_{\mathrm{eff}}$ itself.

The marginalized correlation between $\sum m_{\nu}$ and $N_{\mathrm{eff}}$ is only weakly positive, with a Pearson coefficient of $0.071$, which can be understood from their respective correlations with $\omega_{\mathrm{cdm}}$ and $h$. For $N_{\mathrm{eff}}$, the strong positive partial correlations of $0.979$ with $\omega_{\mathrm{cdm}}$ and $0.950$ with $h$ remain nearly as strong after marginalization, with Pearson coefficients of $0.961$ and $0.937$, respectively. Increasing $N_{\mathrm{eff}}$ directly raises the pre-recombination radiation density, requiring larger $\omega_{\mathrm{cdm}}$ to preserve the MR equality scale and favoring larger $h$ to preserve $\theta_{\mathrm{s}}$, so these strong relationships remain apparent in the marginalized posterior. For $\sum m_{\nu}$, by contrast, the strong negative partial correlations of $-0.830$ with $\omega_{\mathrm{cdm}}$ and $-0.891$ with $h$ are reduced to nearly vanishing Pearson correlations of $-0.005$ and $-0.042$, respectively. The shifts toward smaller $\omega_{\mathrm{cdm}}$ and $h$ favored by larger values $\sum m_{\nu}$ can be offset when $N_{\mathrm{eff}}$ increases simultaneously and drives both parameters upward, largely masking the corresponding correlations with $\sum m_{\nu}$.

By removing these degeneracies with the other cosmological parameters, particularly $\omega_{\mathrm{cdm}}$ and $h$, the strong positive partial correlation of $0.857$ between $\sum m_{\nu}$ and $N_{\mathrm{eff}}$ reflects their opposing effects on $\theta_{\mathrm{s}}$. Larger $\sum m_{\nu}$ raises the value of $\theta_{\mathrm{s}}$, whereas increasing $N_{\mathrm{eff}}$ decreases it, allowing larger values of $\sum m_{\nu}$ and $N_{\mathrm{eff}}$ to compensate for one another. In this analysis, we vary $T_{\nu}$ directly, so increasing the value of $N_{\mathrm{eff}}$ corresponds to a larger relic-neutrino abundance. The strong positive partial correlation between $\sum m_{\nu}$ and $N_{\mathrm{eff}}$ therefore shows that heavier neutrinos can be accommodated by a larger, rather than smaller, neutrino abundance.

\section{Prior-volume effects}
\label{sec:prior_volume_effects}

In \cref{subsec:numerical_implementation}, we describe our two sampling methods for performing MCMC analyses with the NO and IO. In the fixed approach, the sampled prior is flat in $\sum m_{\nu}$, while in the marginalized approach, the sampled prior is flat in $m_0$. The nonlinear mapping between $m_0$ and $\sum m_{\nu}$ induces different priors in the two methods from conservation of probability under a change of variables, $p(\sum m_{\nu}) \mathrm{d}(\sum m_{\nu}) = p(m_0) \mathrm{d}m_0$. The difference between the two prior choices is governed by the Jacobian,
\begin{equation}
    J \equiv \frac{\mathrm{d}(\sum m_{\nu})}{\mathrm{d}m_0}.
    \label{eq:jacobian_m_tot_m_0}
\end{equation}
We therefore write $\sum m_{\nu}$ as a function of $m_0$ and the mass splittings for the NO and IO.  Following the NuFIT convention~\cite{Gonzalez-Garcia:2014bfa}, the solar splitting is defined identically in both orderings, $\Delta m_{\mathrm{sol}}^2 \equiv m_2^2 - m_1^2$, while the atmospheric splitting is defined as $\Delta m_{\mathrm{atm}}^2 \equiv m_3^2 - m_1^2$ in the NO and $\Delta m_{\mathrm{atm}}^2 \equiv m_3^2 - m_2^2$ in the IO. For the NO, the individual mass eigenstates are then given by
\begin{subequations}
\label{eq:masses_no}
\begin{align}
    m_1 &= m_0,
    \label{eq:m_1_no}\\
    m_2 &= \sqrt{m_0^2 + \Delta m_{\mathrm{sol}}^2},
    \label{eq:m_2_no}\\
    m_3 &= \sqrt{m_0^2 + \Delta m_{\mathrm{atm}}^2},
    \label{eq:m_3_no}
\end{align}
\end{subequations}
where $\Delta m_{\mathrm{atm}}^2 > 0$. For the IO, where we instead have $\Delta m_{\mathrm{atm}}^2 < 0$, the individual mass eigenstates are
\begin{subequations}
\label{eq:masses_io}
\begin{align}
    m_3 &= m_0,
    \label{eq:m_3_io}\\
    m_1 &= \sqrt{m_0^2 - \Delta m_{\mathrm{atm}}^2 - \Delta m_{\mathrm{sol}}^2},
    \label{eq:m_1_io}\\
    m_2 &= \sqrt{m_0^2 - \Delta m_{\mathrm{atm}}^2}.
    \label{eq:m_2_io}
\end{align}
\end{subequations}

\begin{figure}
    \includegraphics[width = \columnwidth]{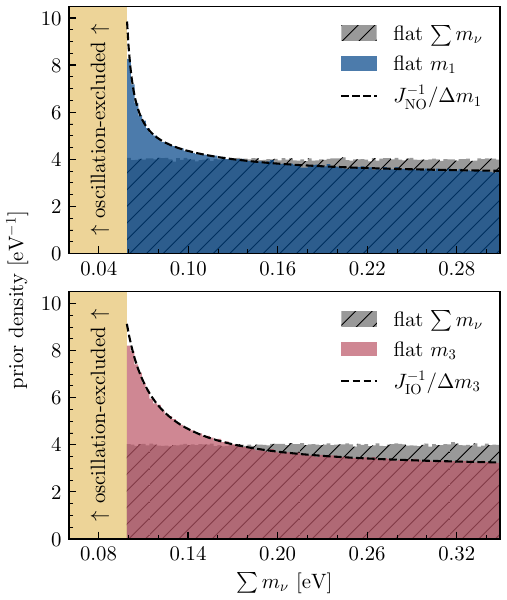}
    \caption{Prior density in $\sum m_{\nu}$ for our two neutrino-mass sampling prescriptions. Gray hatched regions show the fixed method, in which $\sum m_{\nu}$ is sampled with a flat prior and $m_0$ is derived. Colored histograms depict the marginalized method, in which $m_0$ is sampled with a flat prior, the mass splittings are sampled from Gaussian likelihoods, and $\sum m_{\nu}$ is derived. The upper and lower panels use the NO and IO, respectively. We plot \cref{eq:m_tot_prior_distribution} with dashed black curves, showing that the enhancement at low $\sum m_{\nu}$ is driven by the change of variables, with trivial corrections from the splitting uncertainties.}
    \label{fig:m_tot_prior_density}
\end{figure}

In \cref{fig:m_tot_prior_density}, we depict the effects of a flat $m_0$ prior on the induced $\sum m_{\nu}$ prior. To reproduce the fixed approach, samples are drawn uniformly in $\sum m_{\nu}$, producing the constant prior density shown by the gray hatched regions. To reproduce the marginalized approach, $m_0$ is drawn from a flat prior and $\Delta m_{\mathrm{sol}}^2$ and $\Delta m_{\mathrm{atm}}^2$ are drawn from Gaussian priors. We compute $\sum m_{\nu}$ for each sample and construct the colored histograms, showing that a flat $m_0$ prior induces more prior density near the minimum $\sum m_{\nu}$ than a prior that is flat in $\sum m_{\nu}$. This explains why the marginalized method leads to a tighter bound on $\sum m_{\nu}$.

\begin{figure}
    \includegraphics[width = \columnwidth]{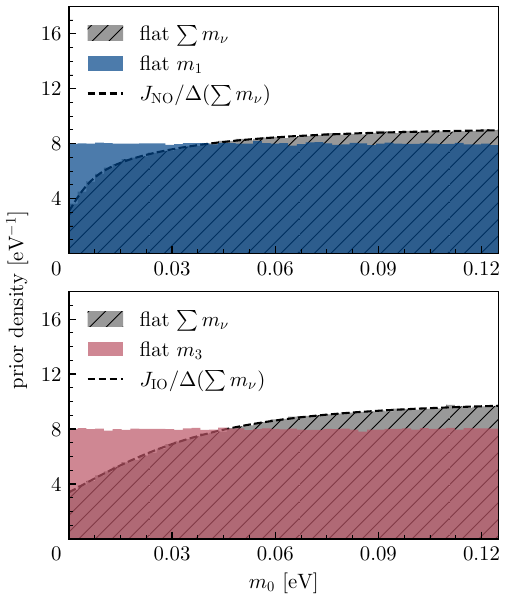}
    \caption{Prior density in $m_0$ for our two neutrino-mass sampling prescriptions. Gray hatched regions show the fixed method, in which $\sum m_{\nu}$ is sampled with a flat prior and $m_0$ is derived. Colored histograms depict the marginalized method, in which $m_0$ is sampled with a flat prior, the mass splittings are sampled from Gaussian likelihoods, and $\sum m_{\nu}$ is derived. The upper and lower panels use the NO and IO, respectively. We plot \cref{eq:m_0_prior_distribution} with dashed black curves, showing that the enhancement at low $m_0$ is due to the change of variables.}
    \label{fig:m_0_prior_density}
\end{figure}

Two subdominant effects alter the $\sum m_{\nu}$ distributions generated from flat $m_0$ priors. First, $\sum m_{\nu}$ depends on the square roots of the mass splittings, so the Gaussian distributions in $\Delta m_{\mathrm{sol}}^2$ and $\Delta m_{\mathrm{atm}}^2$ induce an asymmetric $\sum m_{\nu}$ distribution. Drawing $\Delta m_{\mathrm{sol}}^2$ and $\Delta m_{\mathrm{atm}}^2$ from Gaussian priors also allows a small spread of $\sum m_{\nu}$ values for a given $m_0$. Both features are much smaller than the Jacobian effect, which induces a distribution in $\sum m_{\nu}$ of
\begin{equation}
    p({\textstyle \sum} m_{\nu}) = \frac{J^{-1}}{\Delta m_0},
    \label{eq:m_tot_prior_distribution}
\end{equation}
where $\Delta m_0 \equiv m_{0, \, \max} - m_{0, \, \min}$ is the width of the $m_0$ range sampled in \cref{fig:m_tot_prior_density}. For each ordering, the limits on $m_0$ are obtained by inverting the fixed-splitting relation for $\sum m_{\nu}$ at the endpoints of the corresponding flat $\sum m_{\nu}$ prior. We plot the induced prior distribution for fixed splittings with dashed black curves, and their close agreement with the hatched histograms in gray indicates that the additional spread and asymmetry from the splitting uncertainties are negligible.

The nonlinear mapping between $m_0$ and $\sum m_{\nu}$ also produces a prior-volume effect in $m_0$, as shown in \cref{fig:m_0_prior_density}. In the marginalized method, $m_0$ is sampled with a flat prior. To cover the same $m_0$ range in the fixed method, we compute $\sum m_{\nu}$ at the endpoints of the $m_0$ prior using fixed splittings, and sample uniformly over this range with width $\Delta(\sum m_{\nu}) \equiv (\sum m_{\nu})_{\max} - (\sum m_{\nu})_{\min}$. The Jacobian then induces a distribution in $m_0$ of
\begin{equation}
    p(m_0) = \frac{J}{\Delta({\textstyle \sum} m_{\nu})}.
    \label{eq:m_0_prior_distribution}
\end{equation}
As seen in \cref{fig:m_0_prior_density}, the fixed approach places more prior weight in the high-$m_0$ region than a prior that is flat in $m_0$. This change-of-variables effect explains why the marginalized approach gives tighter upper bounds on $m_0$.

\section{Additional MCMC results}
\label{sec:additional_MCMC_results}

We show marginalized posteriors for all cosmological and decay parameters under the DO in \cref{fig:triangle_all_data}, including the results for the \textit{Planck}+DESI+lens(PR3) dataset. The corresponding constraints for this analysis are presented in \cref{tab:planck_desi_pr3_do}, and the neutrino-mass bound is relaxed from $\sum m_{\nu} < \qty{0.0766}{\electronvolt}$~(95\%~CrI;~DO) for $\Lambda$CDM to $\sum m_{\nu} < \qty{0.0824}{\electronvolt}$~(95\%~CrI;~DO) for $Y$ decay. The green dashed contours for the \textit{Planck}+DESI+lens(PR3) dataset are nearly indistinguishable from the \textit{Planck}+DESI results in solid yellow, indicating that the gain in constraining power from the \textit{Planck} PR3 lensing likelihood is minimal. In comparison with the \textit{Planck}+DESI constraints in \cref{tab:all_datasets_do}, adding PR3 lensing slightly tightens the neutrino-mass bound in $\Lambda$CDM but leaves the $Y$-decay limit essentially unchanged, suggesting that the additional lensing information is largely degenerate with the extra parameter freedom introduced by the $Y$-decay scenario.

\begin{table}
    \caption{Marginalized constraints on the decay and cosmological parameters under the DO approximation and the \textit{Planck}+DESI+lens(PR3)+SD+D/H dataset, quoted at 95\% credibility. For quantities common to both models, the upper entry in regular text shows the $\Lambda$CDM result, while the lower entry in boldface gives the $Y$-decay constraint. The SD and D/H likelihoods are omitted in the $\Lambda$CDM fit.}
    \label{tab:planck_desi_pr3_do}
    \begin{ruledtabular}
    \begin{tabular}{lc}
    
    $f_{\gamma}$
    &
    $\cdots$\\
    \addlinespace[1.5 pt]
    
    $\log_{10}(\Gamma_Y / \unit{\per\second})$
    &
    $\mathbf{> -6.15}$\\
    \addlinespace[1.5 pt]
    
    $R_{\Gamma}$
    &
    $\mathbf{< 0.0252}$\\
    \midrule
    
    $\sum m_{\nu} \, [\unit{\electronvolt}]$
    &
    \begin{tabular}{c}
        $< 0.0766$\\
        $\mathbf{< 0.0824}$
    \end{tabular}\\
    \addlinespace[1.5 pt]
    
    $100 \omega_{\mathrm{b}}$
    &
    \begin{tabular}{c}
        $\num{2.252(0.025:0.025)}$\\
        $\mathbf{\num{2.250(0.026:0.026)}}$
    \end{tabular}\\
    \addlinespace[1.5 pt]
    
    $\omega_{\mathrm{cdm}}$
    &
    \begin{tabular}{c}
        $\num{0.1180(0.0013:0.0013)}$\\
        $\mathbf{\num{0.1190(0.0038:0.0030)}}$
    \end{tabular}\\
    \addlinespace[1.5 pt]
    
    $h$
    &
    \begin{tabular}{c}
        $\num{0.6860(0.0060:0.0061)}$\\
        $\mathbf{\num{0.686(0.012:0.011)}}$
    \end{tabular}\\
    \addlinespace[1.5 pt]
    
    $\ln(10^{10} A_{\mathrm{s}})$
    &
    \begin{tabular}{c}
        $\num{3.047(0.030:0.027)}$\\
        $\mathbf{\num{3.048(0.035:0.033)}}$
    \end{tabular}\\
    \addlinespace[1.5 pt]
    
    $n_{\mathrm{s}}$
    &
    \begin{tabular}{c}
        $\num{0.9699(0.0066:0.0066)}$\\
        $\mathbf{\num{0.9711(0.0089:0.0079)}}$
    \end{tabular}\\
    \addlinespace[1.5 pt]
    
    $\tau_{\mathrm{reio}}$
    &
    \begin{tabular}{c}
        $\num{0.058(0.015:0.014)}$\\
        $\mathbf{\num{0.057(0.017:0.015)}}$
    \end{tabular}
    \end{tabular}
    \end{ruledtabular}
\end{table}

In \cref{fig:triangle_entropy_injection_update}, we present updated constraints on the $Y$-decay scenario with fixed $\sum m_{\nu}$, following Ref.~\cite{Sobotka:2022vrr} in assuming one massive neutrino species with $m_{\nu} = \qty{0.06}{\electronvolt}$. We extend the analysis of Ref.~\cite{Sobotka:2022vrr} by including DESI DR2 BAO data and CMB lensing measurements from \textit{Planck} PR4 and ACT. The inclusion of these datasets introduces a preference for $f_{\gamma} \lesssim 0.59$ that was not present in Ref.~\cite{Sobotka:2022vrr}, mirroring the behavior found in \cref{sec:results} when marginalizing over $\sum m_{\nu}$. The $f_{\gamma} \lesssim 0.59$ region corresponds to decays that increase $N_{\mathrm{eff}}$ relative to $\Lambda$CDM, which are favored by DESI because they correspond to a smaller $\Omega_{\mathrm{m}}$. For our fixed-$\sum m_{\nu}$ analysis, we obtain constraints on the decay parameters of
\begin{subequations}
\label{eq:results_entropy_injection_update}
\begin{align}
    \log_{10}(\Gamma_Y / \unit{\per\second}) &> -6.15,
    \label{eq:decay_rate_entropy_injection_update}\\
    R_{\Gamma} &< 0.0261,
    \label{eq:max_ratio_entropy_injection_update}
\end{align}
\end{subequations}
at 95\% credibility. The marginalized posterior does not yield a 95\% credible interval for $f_{\gamma}$. Relative to the fixed-$\sum m_{\nu}$ analysis of Ref.~\cite{Sobotka:2022vrr}, the lower bound on the decay rate that we obtain is unchanged, while the upper bound on $R_{\Gamma}$ is slightly weakened from $R_{\Gamma} < 0.0235$~(95\%~CrI). The increased posterior weight at $f_{\gamma} \lesssim 0.59$ favors the region in which larger decay amplitudes are allowed, slightly weakening the marginalized bound on $R_{\Gamma}$.

As the fixed-$\sum m_{\nu}$ analysis assumes one massive neutrino species with $m_{\nu} = \qty{0.06}{\electronvolt}$, our NO run with fixed mass splittings offers an apt comparison by imposing a prior of $\sum m_{\nu} \, [\unit{\electronvolt}] \in [0.059, \infty)$. As shown in \cref{tab:full_dataset_do_no_io}, this NO analysis yields $\log_{10}(\Gamma_Y / \unit{\per\second}) > -6.15$~(95\%~CrI) and $R_{\Gamma} < 0.0271$~(95\%~CrI), which are highly similar to the constraints of $\log_{10}(\Gamma_Y / \unit{\per\second}) > -6.15$~(95\%~CrI) and $R_{\Gamma} < 0.0261$~(95\%~CrI) obtained for a single massive neutrino species with $m_{\nu} = \qty{0.06}{\electronvolt}$. These results indicate that allowing $\sum m_{\nu}$ to vary above the minimum mass in the NO has little effect on the decay parameters.

Across the DO, NO, and IO, changing the minimum $\sum m_{\nu}$ does affect the  $R_{\Gamma}$ constraint. On the $f_{\gamma} \lesssim 0.59$ branch preferred by DESI, larger $R_{\Gamma}$ increases $N_{\mathrm{eff}}$ and favors larger $h$ and $\omega_{\mathrm{cdm}}$. Increasing $\sum m_{\nu}$ shifts both $h$ and $\omega_{\mathrm{cdm}}$ in the opposite direction and can partially compensate the changes in these parameters when increasing $R_{\Gamma}$ for decays with $f_{\gamma} \lesssim 0.59$. For fixed splittings, the $R_{\Gamma}$ bound consistently loosens from $R_{\Gamma} < 0.0232$~(95\%~CrI) in the DO to $R_{\Gamma} < 0.0271$~(95\%~CrI) in the NO and $R_{\Gamma}< 0.0297$~(95\%~CrI) in the IO as the minimum allowed value of $\sum m_{\nu}$ increases.

{\renewcommand{\dblfloatpagefraction}{0.5}
\begin{figure*}[p]
    \includegraphics[width = \textwidth]{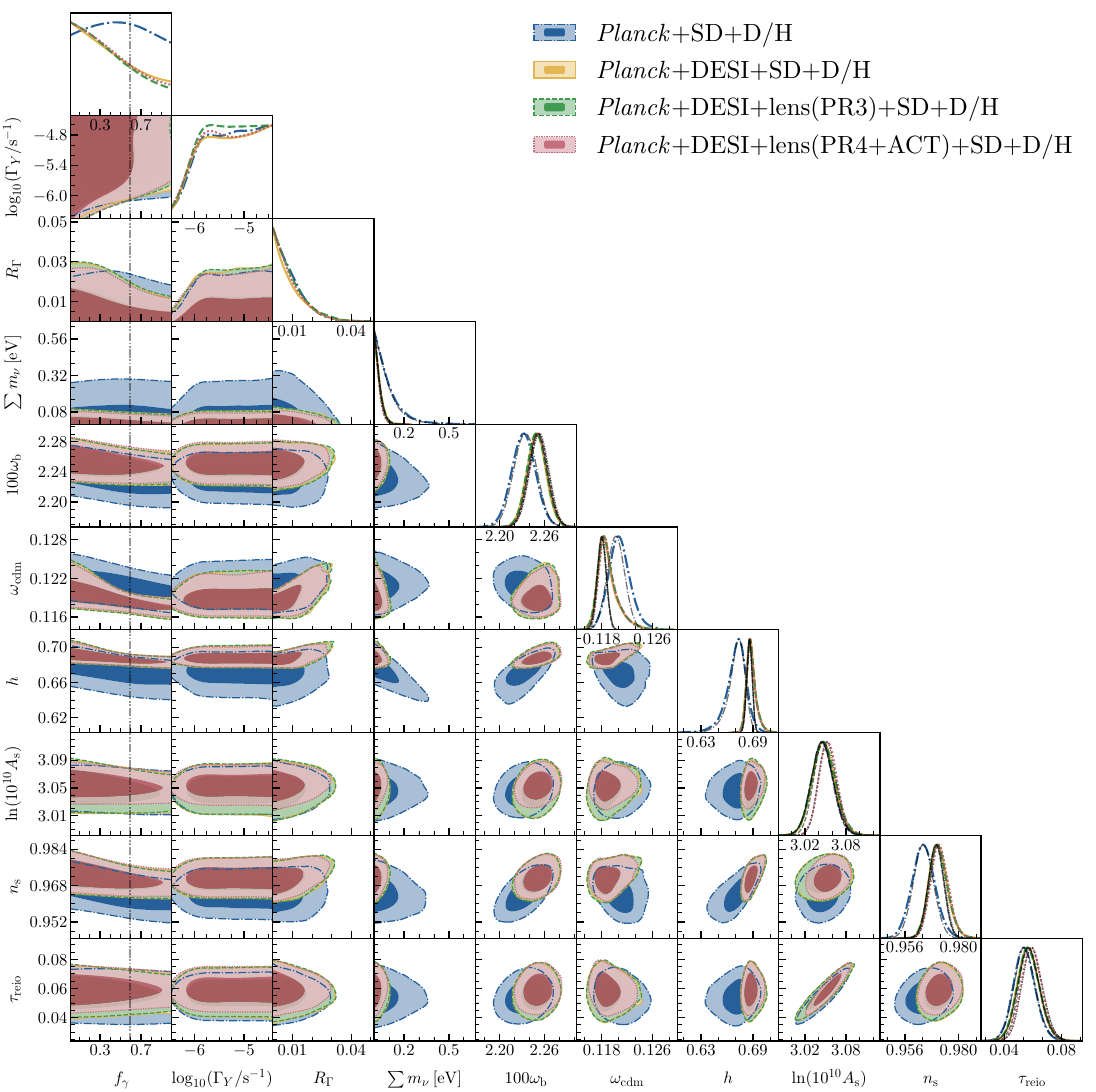}
    \caption{Extended version of \cref{fig:triangle_principal_data} showing the full set of decay and cosmological parameters for the DO analysis, including the \textit{Planck}+DESI+lens(PR3) dataset. We plot the one-dimensional posteriors for $\Lambda$CDM using the same line style but in gray and with a narrower line width than their $Y$-decay counterparts. The vertical line in the first column denotes $f_{\gamma} = 0.5913$. $\Lambda$CDM runs use the same datasets as indicated in the legend, but with the SD and D/H likelihoods omitted.}
    \label{fig:triangle_all_data}
\end{figure*}

\begin{figure*}[p]
    \includegraphics[width = \textwidth]{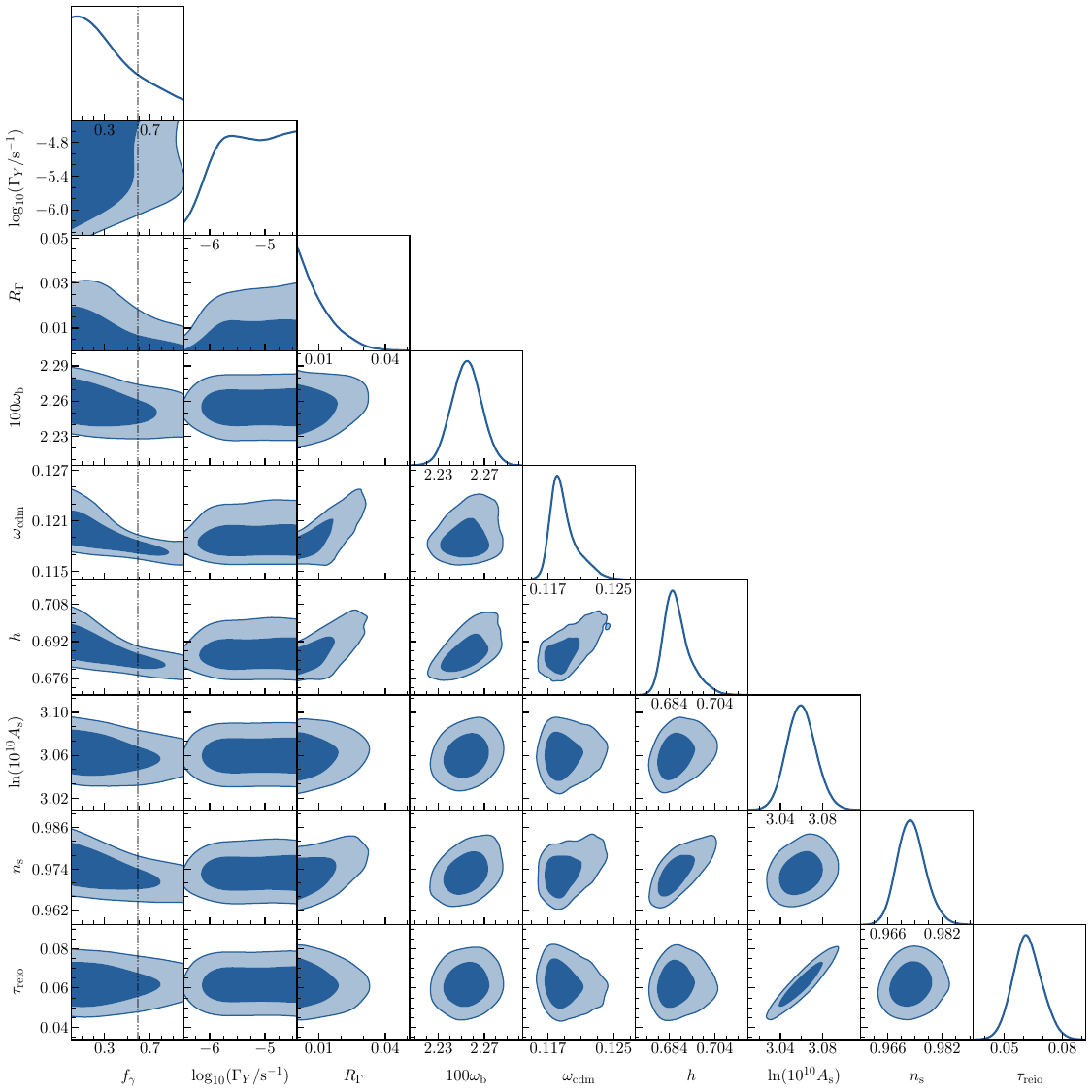}
    \caption{Updated constraints on entropy injection scenarios from Ref.~\cite{Sobotka:2022vrr}, assuming one massive neutrino species with $m_{\nu} = \qty{0.06}{\electronvolt}$. In addition to the \textit{Planck} PR3 TT,TE,EE, SD, and D/H likelihoods used in that work, we include the DESI DR2 BAO sample and CMB lensing reconstructions from \textit{Planck} PR4 \texttt{NPIPE} and ACT DR6. The vertical line in the first column denotes $f_{\gamma} = 0.5913$. We obtain a new preference for $f_{\gamma} \lesssim 0.59$, rather than the slight peak at $f_{\gamma} \approx 0.59$ found in Ref.~\cite{Sobotka:2022vrr} corresponding to decays that preserve $N_{\mathrm{eff}} \simeq 3.04$. As shown in \cref{sec:results}, decays with $f_{\gamma}>0.5913$ result in $N_{\mathrm{eff}} < 3.044$ and favor smaller $h$, which is strongly disfavored by the DESI data newly included in our updated analysis.}
    \label{fig:triangle_entropy_injection_update}
\end{figure*}
\pagebreak}

\bibliographystyle{apsrev4-2}
\bibliography{references}

\end{document}